\documentclass[twocolumn,prb, showpacs, superscriptaddress]{revtex4}
\usepackage{amssymb}
\usepackage{amsmath}
\usepackage{bbm}
\usepackage{graphicx}

\begin{document}

\author{Philipp Werner}
\affiliation{Department of Physics, University of Fribourg, 1700 Fribourg, Switzerland}
\author{Martin Eckstein}
\affiliation{Max Planck Research Department for Structural Dynamics, University of Hamburg-CFEL, Hamburg, Germany}
\title{Phonon-enhanced relaxation and excitation in the Holstein-Hubbard model}

\date{\today}

\hyphenation{}

\begin{abstract}
We study quenches of the interaction and electron-phonon coupling parameter in the Hubbard-Holstein model, using nonequilibrium dynamical mean field theory. The calculations are based on a generalized Lang-Firsov scheme for time-dependent interactions or externally driven phonons, and an approximate strong-coupling impurity solver. The interaction quench calculations reveal the phonon-assisted decay of excess doublons, while the quenches of the electron-phonon coupling lead to persistent oscillations of the phonons and to a phonon-enhanced doublon production.
\end{abstract}


\maketitle

\section{Introduction}

Pump-probe spectroscopy provides a powerful tool to explore the nonequilibrium properties of correlated solids on the relevant time-scales for the electron and phonon dynamics. Many materials of present day interest, such as high-$T_c$ cuprates and organic superconductors, exhibit strong electron-electron and sizeable electron-phonon couplings, and in equilibrium it is difficult to disentangle the effect of these two types of interactions. Time-resolved experiments which selectively excite either the electrons or phonons can provide new insights and even lead to metastable states of matter with interesting properties. For example, in the insulating charge-density wave compound 1$T$-TaS$_{2}$, a purely electronically driven insulator-to-metal transition has been found in time-resolved photo-emission experiments.\cite{Perfetti2006, Perfetti2008} 
Optical pump-probe spectroscopy was used to estimate electron-phonon coupling strengths in high-$T_c$ superconductors.\cite{Gadermaier2010, Gadermaier2012, Conte2012} 
Squeezed phonon states, resulting from a pulse-induced softening of certain phonon modes in bismuth, have been measured with time-resolved $X$-ray pump-probe spectroscopy.\cite{Johnson2009} Most interestingly, the selective excitation of apical oxygen modes in certain cuprate materials using THz pulses, has been shown to induce a transient superconducting state above the equilibrium $T_c$.\cite{Fausti2010, Kaiser2013} 

Motivated by these experimental developments, several theoretical works 
have recently addressed
the nonequilibrium dynamics of electron-phonon coupled systems. For example, 
a mapping from a time-dependent Boltzmann equation onto a Schr\"odinger type equation was used to predict electron relaxation times in metallic systems,\cite{Kabanov2008} under the assumption that the phonons remain in a state of thermal equilibrium. 
Ref.~\onlinecite{Yonemitsu2009} studied the energy transfer from the electronic system to lattice vibrations in a one-dimensional photoexcited Mott insulator, based on a numerical integration of the time-dependent Schr\"odigner equation and a classical phonon approximation. A quantum mechanical treatment of the dynamics and nonlinear transport characteristics of one or two polarons in the Holstein model was presented in Refs.~\onlinecite{Vidmar2011,Golez2012a, Golez2012b}. These calculations employed a time-dependent Lanczos scheme in a variational Hilbert space.\cite{Bonca1999} Pump excitations of the two-dimensional Holstein model were investigated using time-dependent lattice perturbation theory (Migdal approximation).\cite{Kemper2012} 

Few methods exist which can handle lattice models in $d>1$ and in the particularly challenging regime of strong electron-electron and 
electron-phonon interaction. The two-temperature model,\cite{Allen1987} which assumes that the electronic relaxation is fast compared to the timescale of phonons, is certainly inadequate in this case, since it has been demonstrated that the relaxation time in purely electronic systems with a gap can be much longer than typical phonon oscillation times.\cite{Eckstein2010pump} It is therefore important to develop a formalism which can describe the feedback of the quantum phonons on the electronic relaxation process and the effect of the nonequilibrium state of the electrons on the evolution of the phonons. 

In equilibrium, the Holstein-Hubbard model, which describes a coupling to local (Einstein) phonons, has been widely used to study the interplay of electron-electron and electron-phonon interactions.\cite{Freericks1995, Koller2004, Tezuka2005, Werner2007, Bauer2010} It captures a variety of physics, including the Mott metal-insulator transition and phonon-driven polaron and bipolaron formation, but cannot be solved exactly in the general case. A numerical investigation in the high-dimensional limit is possible within the dynamical mean field (DMFT) framework,\cite{Georges1996} and this formalism can also be applied, with rather straightforward generalizations, to nonequilibrium situations.\cite{Schmidt2002, Freericks2006} Over the last few years, nonequilibrium DMFT has been used to study relaxation phenomena,\cite{Eckstein2009, Eckstein2010} photo-doping\cite{Eckstein2013} and symmetry-breaking transitions in the Hubbard model.\cite{Werner2012, Tsuji2012} Here, we extend the nonequilibrium DMFT method to the Holstein-Hubbard model, to explore interaction quenches and phonon-coupling quenches in the regime 
of
strong electron-electron and electron-phonon coupling.  

The outline of this paper is as follows. In Sec.~\ref{model} we discuss the model and its solution based on a strong-coupling (hybridization expansion) approach, as well as approximate strong-coupling impurity solvers. In Sec.~\ref{results} we apply the formalism to interaction and phonon-coupling quenches and explore the time-evolution of the double occupancy and spectral function. Sec.~\ref{summary} is a summary and conclusion.

\section{Model and method \label{model}}

\subsection{Hybridization expansion for the Holstein-Hubbard model}

A simple model for strongly correlated materials is the Hubbard model, 
which describes the hopping of electrons between neighboring lattice sites and an on-site repulsion 
between electrons of opposite spin. A local coupling to dispersionless phonons can be included
along the lines proposed by Holstein,\cite{Holstein1959} leading to the so-called Holstein-Hubbard model, 
\begin{align}
H(t)=&-\sum_{i,\delta,\sigma}v c^\dagger_{i+\delta,\sigma}c_{i,\sigma}+\omega_0(t)\sum_i b^\dagger_i b_i\nonumber\\
&+\sum_i \left[U(t)n_{i,\uparrow}n_{i,\downarrow}-\mu(t)(n_{i,\uparrow}+n_{i,\downarrow})\right]\nonumber\\
&+\sum_i \left[\lambda(t)(n_{i,\uparrow}+n_{i,\downarrow}-1)+\omega_0 F(t)\right] (b^\dagger_i+b_i).
\label{H}
\end{align}
Here, $U$ denotes the on-site repulsion, $\mu$ the chemical potential of the electrons with creation operators $c^\dagger_\sigma$ and density operators $n_\sigma$, $b^\dagger$ the creation operator for Einstein phonons of frequency $\omega_0$, and the electron-phonon coupling is $\lambda$. 
The hopping matrix element is denoted by $v$. 
Because we will consider only situations where no external electromagnetic field is coupled directly to the electrons,  the only property of the hopping 
which is relevant in a DMFT calculation is the density of states  ${\cal D}(\omega)=\sum_p \delta(\omega-\varepsilon_p)$, where the energy 
dispersion $\varepsilon_p$ is defined as the Fourier transform of the hopping matrix.
Initially, at time $t=0$, the system is assumed to be in an equilibrium state corresponding to the interaction parameters $U(0)$ and $\lambda(0)$, and 
phonon frequency $\omega_0(0)$.  A nontrivial time-evolution may then be triggered either by an {\it interaction quench} (rapid change of $U(t)$), 
a {\it phonon-coupling quench} (rapid change of $\lambda(t)$), or a {\it phonon-frequency quench} (rapid change of $\omega_0(t)$). To describe 
the external driving of the phonons via a dipole coupling to a time-dependent electric field, we also added the term proportional to $F(t)$ (assuming $F(0)=0$). 
A coupling of the external electromagnetic field to the electrons will not be considered in this work. However, the extension of the formalism 
would be straightforward, because one would  have to modify only the DMFT self-consistency, in the same way as it is done in the  Hubbard model without 
phonons.\cite{Freericks2006}

We compute the time evolution of model (\ref{H}) using the single-site dynamical mean field (DMFT) approximation,\cite{Georges1996} which reduces
the problem to the solution of a quantum impurity model 
(one interacting site coupled to a bath
of noninteracting conduction electrons) combined with a self-consistency condition. The DMFT formalism can be applied to nonequilibrium 
problems,\cite{Schmidt2002, Freericks2006} by extending the imaginary-time interval to an $L$-shaped Kadanoff-Baym contour $\mathcal{C}$ (see Fig.~\ref{illustration}). 

The quantum impurity model which must be solved in DMFT calculations of the Holstein-Hubbard model can be 
specified by the Hamiltonian
\begin{equation}
H_\text{QI}(t) = H_\text{loc}(t)+H_\text{hyb}(t)+H_\text{bath}(t),
\end{equation}
where the local term is 
\begin{eqnarray}
H_\text{loc}(t) &=& U(t) n_\uparrow n_\downarrow-\mu(t)(n_\uparrow+n_\downarrow)\nonumber\\
&& +\left[\lambda(t) (n_\uparrow+n_\downarrow-1)+\omega_0 F(t)\right] (b^\dagger+b)\nonumber\\
&& + \omega_0(t)b^\dagger b,
\label{H_loc}
\end{eqnarray}
and the impurity-bath mixing and bath Hamiltonians are
\begin{eqnarray}
H_\text{hyb}(t) &=& \sum_{p, \sigma} V_{p, \sigma}(t) c^\dagger_\sigma a_{p, \sigma} + V^*_{p, \sigma}(t) c_\sigma a^\dagger_{p, \sigma},\label{H_hyb}\\
H_\text{bath}(t) &=& \sum_{p, \sigma} \epsilon_p(t) a^\dagger_{p, \sigma} a_{p, \sigma}.
\end{eqnarray}
The parameters $V_{p, \sigma}$ and $\epsilon_p$ are in general time-dependent and determined by the DMFT self-consistency equation.
They enter the DMFT formalism only via the hybridization function $\Lambda_\sigma$, which can be obtained 
directly from the impurity Green's function $G_\sigma(t,t')$, as explained below (Eq.~(\ref{bethe})). 

A numerically exact solution of the impurity model is possible, in principle, using the continuous-time Monte Carlo technique.\cite{Gull2011, Werner2009} For the equilibrium Holstein-Hubbard model, the hybridization expansion approach,\cite{Werner2006} combined with a Lang-Firsov decoupling of the electron-phonon term,\cite{Lang1962, Werner2007} allows very efficient simulations. Here, we will explain how this exact approach can be extended to time-dependent couplings and forces. For the actual simulations, we will then resort to an approximate strong-coupling impurity solver based on the non-crossing approximation.\cite{Keiter1971, Eckstein2010nca} 
  
The hybridization expansion on the Kadanoff-Baym contour $\mathcal{C}$ is based on a perturbation
expansion in $H_\text{hyb}$ defined in Eq.~(\ref{H_hyb}).
After tracing out the bath states $a_{p, \sigma}$, the complex weight of a Monte Carlo configuration corresponding to a perturbation order $n$ ($n$ creation operators $c^\dagger_\sigma(\tau_\sigma)$ and $n$ annihilation operators $c_\sigma(\tau'_\sigma)$) can be expressed as\cite{Werner2006, Werner2007, Werner2009}
\begin{eqnarray}
&&w(\{O_i(t_i)\}) = \text{Tr}_{c,b} \Big[  T_\mathcal{C} e^{-i\int_\mathcal{C} dt H_\text{loc}(t)} O_{2n}(t_{2n})\ldots\nonumber\\
&&\ldots O_{1} (t_{1}) \Big] (-i)^{2n}dt_1\ldots dt_{2n}\prod_\sigma (\det M_\sigma^{-1}), \hspace{5mm}
\label{weight}
\end{eqnarray}
where the $O_i(t_i)$ are the creation and annihilation operators for spin up and down
electrons on the impurity site and the $t_i$ are times on the contour $\mathcal{C}$ 
(the $dt_i$ contain factors $+1$, $-1$ or $-i$, depending on the position of $t_i$ on the contour).
The matrix elements $M_\sigma^{-1}(i,j)=\Lambda_\sigma(t'_{\sigma,i},t_{\sigma,j})$ are given by the hybridization function $\Lambda_\sigma$ (which itself is determined by the time-dependent parameters $V_{p,\sigma}$  and $\epsilon_p$). In practice, $\Lambda_\sigma$ is obtained directly from the impurity Green's function $G_\sigma(t,t')$. In the simple case of a semi-circular density of states of bandwidth $4v$, the relation reads 
\begin{equation}
\Lambda_\sigma(t,t')=v^2G_\sigma(t,t').\label{bethe}
\end{equation}

The time evolution operator in the trace is given by $H_\text{loc}$, which includes a time-dependent electron-phonon coupling. Our goal is to evaluate the trace over the phonon states analytically, and in order to do this, we must decouple the electrons and phonons using a suitable unitary transformation. The procedure in equilibrium has been detailed in Ref.~\onlinecite{Werner2007} and employs the standard Lang-Firsov transformation.\cite{Lang1962} In the presence of time-dependent couplings, a more general transformation 
is
needed, and depending on the type of perturbation, we obtain different variants of the hybridization expansion algorithm. In the following subsections, we explicitly discuss the 
algorithms for externally driven phonons, 
or an arbitrary time-dependence of the interaction, the electron-phonon coupling, and the phonon frequency.

\subsection{Externally driven phonons}

\subsubsection{Exact formalism}

We start by considering an impurity Hamiltonian of the form (\ref{H_loc}), but with constant $U(t)=U$, $\lambda(t)=\lambda$ and $\omega_0(t)=\omega_0$. 
Defining the operators $X=(b^\dagger +b)/\sqrt{2}$ and $P=i(b^\dagger -b)/\sqrt{2}$
satisfying $[X, P]=i$, the unitary transformation $\tilde O(t)=W^\dagger(t)OW(t)$ specified by 
\begin{align}
W(t)&=e^{iPX_0(t)},\label{LFtrafo}\\
X_0(t)&=\sqrt{2}\lambda(n_\uparrow+n_\downarrow-1)/\omega_0+\sqrt{2}F(t),\label{x0_F}
\end{align}
shifts $X$ to 
$X-X_0(t)$, 
so that  the transformed Hamiltonian at time $t$,
\begin{align}
\tilde H_\text{loc}(t) &= W^\dagger(t)H_\text{loc} W(t)\nonumber\\
&= -\tilde \mu(\tilde n_\uparrow+\tilde n_\downarrow)+\tilde U \tilde n_\uparrow \tilde n_\downarrow+\frac{\omega_0}{2}(X^2+P^2), 
\label{H_loc_transf}
\end{align}
has no explicit electron-phonon coupling. ${\tilde H}_\text{loc}$ is the local Hamiltonian with chemical potential and interaction strength
shifted as
\begin{eqnarray}
\tilde \mu&=&\mu-\lambda^2/\omega_0+2\lambda F(t),\label{mu_tilde}\\
\tilde U&=&U-2\lambda^2/\omega_0.\label{U_tilde}
\end{eqnarray}
In principle, there is also a time-dependent constant term $-(\lambda/\omega_0-F(t))^2\omega_0$, 
but this should 
not have any effect on the nonequilibrium dynamics (except on the total energy).

The transformed electron creation and annihilation operators are 
\begin{eqnarray}
\tilde c^\dagger_\sigma &=& W^\dagger(t)c^\dagger_\sigma W(t) = e^{\frac{\lambda}{\omega_0}(b^\dagger-b)}c^\dagger_\sigma,\label{eq_cdag}\\
\tilde c_\sigma &=& W^\dagger(t)c_\sigma W(t) = e^{-\frac{\lambda}{\omega_0}(b^\dagger-b)}c_\sigma.
\end{eqnarray}

To investigate the effect of the time-dependence of the Lang-Firsov transformation, 
let us discretize the path-integral in Eq.~(\ref{weight}) with small time-steps $\Delta\tau$ and insert identiy operators $\mathcal{I}=W^\dagger(t)W(t)$ at each time step. 
The zero-order expression becomes a product of time-evolution operators $\ldots e^{-i\Delta t H_\text{loc}(t+\Delta t)}e^{-i\Delta t H_\text{loc}(t)} \ldots$, which
after the insertion of the identity operators can be regrouped as
\begin{align}
&\ldots e^{-i\Delta t H_\text{loc}(t+\Delta t)}W(t+\Delta t)\Big]\nonumber\\
&\hspace{10mm}\times \Big[W^\dagger(t+\Delta t) W(t)W^\dagger(t) e^{-i\Delta t H_\text{loc}(t)}W(t)\Big]\nonumber\\
&\hspace{20mm}\times\Big[W^\dagger(t)W(t-\Delta t)W^\dagger(t-\Delta t) \ldots .\nonumber
\end{align}
Since $W^\dagger(t+\Delta t)W(t)=e^{-iPX_0(t+\Delta t)}e^{iPX_0(t)}=e^{-iPX'_0(t)\Delta t}$ and $X'_0(t)=\sqrt{2}F'(t)$ we obtain, in addition to the Lang-Firsov transformed $\tilde H_\text{loc}(t)$, a term $PX'_0(t)=P\sqrt{2}F'(t)$.

After separating the bosonic from the fermionic operators,
the weight (\ref{weight}) can thus be written as a product
\begin{align}
w(\{O_i(t_i)\})=&w_b(\{O_i(t_i)\})\tilde w_\text{Hubbard}(\{O_i(t_i)\}),
\label{weight1}
\end{align}
where $\tilde w_\text{Hubbard}$ is the weight of a corresponding configuration in the 
Hubbard (Anderson) impurity model with parameters modified according to Eqs.~(\ref{mu_tilde}) 
and (\ref{U_tilde}), and $w_b$ is the bosonic expectation value
\begin{align}
w_b&=\frac{1}{Z_b}\text{Tr}_{b} \Big[  T_\mathcal{C} e^{-i\int_\mathcal{C} dt H_b(t)} O^b_{2n}(t_{2n})\ldots O^b_{1} (t_{1}) \Big].
\end{align}
Here, $O^b(t)=e^{\pm (\lambda/\omega_0)(b^\dagger(t)-b(t))}$ [plus (minus) sign for time-arguments associated with fermionic creation (annihilation) operators] and the time-dependent Hamiltonian is 
\begin{align}
H_b(t)&=\omega_0b^\dagger(t) b(t) + \sqrt{2}F'(t)i\frac{b^\dagger(t)-b(t)}{\sqrt{2}}\nonumber\\
&=\frac{\omega_0}{2}(X^2(t)+P^2(t))+\sqrt{2}F'(t)P(t).
\end{align}
To evaluate $w_b$, we solve the Heisenberg equations
\begin{align}
X'(t) & = i[H_b(t), X(t)] = +\omega_0 P(t) +\sqrt{2}F'(t),\\
P'(t) &= i[H_b(t),P(t)] = -\omega_0 X(t),
\end{align} 
which gives
\begin{align}
X(t)=&X(0)\cos(\omega_0 t)+P(0)\sin(\omega_0 t)\nonumber\\
&+\int_0^t d\bar t \cos(\omega_0(t-\bar t))\sqrt{2}F'(\bar t),\nonumber\\
P(t)=&P(0)\cos(\omega_0 t)-X(0)\sin(\omega_0 t)\nonumber\\
&-\int_0^t d\bar t \sin(\omega_0(t-\bar t))\sqrt{2}F'(\bar t),\nonumber 
\end{align}
with $X(0)=(b^\dagger+b)/\sqrt{2}$, $P(0)=i(b^\dagger-b)/\sqrt{2}$, and thus
\begin{align}
b^\dagger(t)&=\frac{X(t)-iP(t)}{\sqrt{2}}=b^\dagger e^{i\omega_0 t}+\int_0^t d\bar t F'(\bar t)e^{i\omega_0(t-\bar t)},\nonumber\\
b(t)&=\frac{X(t)+iP(t)}{\sqrt{2}}=b e^{-i\omega_0 t}+\int_0^t d\bar t F'(\bar t)e^{-i\omega_0(t-\bar t)}.\nonumber
\end{align}
Introducing the variable $s=\pm\frac{\lambda}{\omega_0}$, we can write the operator $O^b(t)$ as
\begin{align}
O^b(t)&=e^{s(b^\dagger(t)-b(t))}\nonumber\\
&=e^{s(b^\dagger e^{i\omega_0t}-b e^{-i\omega_0t})}e^{2is\int_0^td\bar t F'(\bar t)\sin(\omega_0(t-\bar t))}\nonumber\\
&=O^b_{F=0}(t)e^{2i\omega_0 s\int_0^td\bar t F(\bar t)\cos(\omega_0(t-\bar t))},\label{Ob}
\end{align}
where in the last step, we used the fact that $F(t=0)=0$ to reexpress the integral. $O^b_{F=0}(t)=e^{s(b^\dagger e^{i\omega_0t}-b e^{-i\omega_0t})}$ is the operator in the absence of external driving. It follows immediately that the bosonic factor $w_b$ is of the form
\begin{align}
w_b=& w_b^{F=0}w_b^\text{ext},
\end{align}
where
\begin{align}
w_b^\text{ext}=&\exp\Bigg[2i\omega_0 \sum_{k=1}^{2n} s_k\int_0^{t_k}d\bar t F(\bar t)\cos(\omega_0(t_k-\bar t))\Bigg]
\end{align}
and $w_b^{F=0}$ is given by the same expression as in the equilibrium Holstein-Hubbard formalism\cite{Werner2007} 
\begin{align}
w_b^{F=0}=&\exp\Bigg[-\frac{1}{\sinh(\beta\omega_0/2)}\bigg(\sum_n \frac{s_n^2}{2}\cosh(\beta\omega_0/2)\nonumber\\
&+\sum_{n>m}s_n s_m\cosh((\beta/2-i(t_n-t_m))\omega_0)\bigg)\Bigg].\label{wbF0}
\end{align}
In Eq.~(\ref{wbF0}) it is assumed that the times $t_1<t_2<\ldots < t_{2n}$ are ordered along the contour $\mathcal{C}$. 
\begin{figure}[t]
\begin{center}
\includegraphics[angle=0, width=0.9\columnwidth]{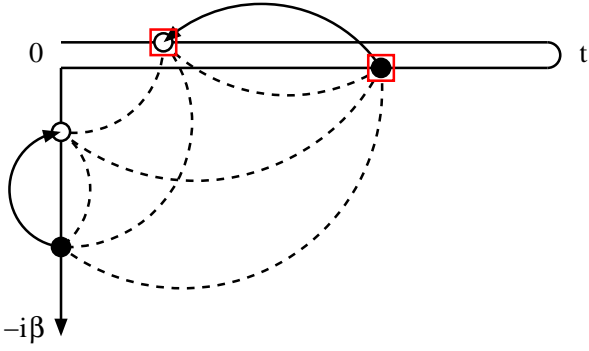}
\caption{Illustration of a strong-coupling diagram on the Kadanoff-Baym contour $\mathcal{C}$ (perturbation order $n=2$). The dashed lines are the retarded interactions between creation (full dots) and annihilation (empty dots) operators, which arise from the coupling to the phonons. The effect of the driving is represented by terms (red boxes), which act locally at the positions of the operators which lie on the real-time branches of the contour.
}
\label{illustration}
\end{center}
\end{figure}
For the total weight (\ref{weight1}) we thus obtain the expression 
\begin{align}
w(\{O_i(t_i)\})=&w_b^{F=0}(\{O_i(t_i)\})w_b^\text{ext}(\{O_i(t_i)\})\nonumber\\
&\times \tilde w_\text{Hubbard}(\{O_i(t_i)\}).
\end{align}
A strong-coupling diagram may thus be represented as sketched in Fig.~\ref{illustration} (example for perturbation order $n=2$). 
The locations of the hybridization operators are marked by full (creation operators) and empty
(annihilation operators) circles on the contour $\mathcal{C}$, which runs from 0 to $t$ along the real axis, back to 0 and then to $-i\beta$ along the imaginary time axis. The weight $w_b^{F=0}(\{O_i(t_i)\})$ can be interpreted as originating from ``interaction lines" between all pairs of operators, with weight
\begin{align}
w^\text{line}&=\exp\Bigg[-\frac{s^<s^>}{\sinh(\beta\omega_0/2)}\big\{\cosh((\beta/2-i(t^>-t^<))\omega_0)\nonumber\\
&\hspace{2cm}-\cosh(\beta\omega_0/2)\big\}\Bigg] \label{eq_wline}
\end{align}
(dahed lines in the figure), while the weight $w_b^\text{ext}(\{O_i(t_i)\})$ can be taken into account by assigning an additional weight  
\begin{align}
w^\text{box}&=
\exp\Bigg[2is\omega_0\int_0^{t} d\bar t F(\bar t)\cos((t-\bar t)\omega_0)\Bigg]\label{wbox}
\end{align} 
to each operator on the real-time branches of the contour (red boxes in the figure). In these formulas, the greater (lesser) signs in the exponents refer to the operators with larger (smaller) time argument on the contour. Pairs of creation and annihilation operators are linked by hybridization lines (solid lines with arrows in the figure). The $n!$ possible ways of connecting creation and annihilations operators by hybridization lines are summed up in the factor $\det M^{-1}$ contained in $\tilde w_\text{Hubbard}$ (see Eq.~(\ref{weight}) and Refs.~\onlinecite{Werner2006,Werner2007}).

\subsubsection{Self-consistency}

The external force appears only in the weight factor $w_b^\text{ext}$ (boxes in Fig.~\ref{illustration}), while $w_b^{F=0}$ (dashed lines) corresponds to the phonon weight for $F=0$. One can show that in a homogeneous system, the ``box" contributions cancel. An easy way to see this is to consider the self-consistency for the semi-circular density of states, $\Lambda = v^2 G$. The diagrams for the Green's function $G$ also have ``box" terms attached to the operators $c^\dagger$ and $c$, but they are complex conjugate to the terms attached to $\Lambda$ (because $G(t,t')\propto \partial \Omega /\partial \Lambda(t',t)$, where $\Omega$ is the grand potential).
Hence, we can simply ignore $w_b^\text{ext}$, perform the DMFT calculation for $F=0$, and multiply the converged $G$ with the appropriate $w^\text{box}$ factors. This however means that the external driving has no effect on local quantities, such as the double occupancy, and that the effect on nonlocal quantities such as Green's functions is trivial in the sense that it does not propagate into the self-consistent calculation.

More generally, we can understand the rather trivial effect of a site-independent driving term in a homogeneous system as follows: suppose that we expand the path integral expressions for the time-dependent double occupancy or Green's function of the lattice model in the hopping terms $t_{ij}c^\dagger_i c_j$. After the decoupling of the electron-phonon interaction on each site by a Lang-Firsov transformation of the form (\ref{LFtrafo}), (\ref{x0_F}), each fermionic operator in these hopping terms gets multiplied by a factor which is identical to Eq.~(\ref{Ob}), apart from a site index. Now, since $s=\pm \frac{\lambda}{\omega_0}$ for creation/annihilation operators, and $F(t)$ is supposed to be site-independent, the $F$-dependent exponential factors will cancel for each hopping term. In the case of a local observable, such as the double occupancy, the remaining expression is identical to the expansion one would get for the system with $F=0$, apart from a trivial shift in the total energy which comes from the $F$-dependent shift of the Lang-Firsov transformed chemical potential. In the case of a Green's function, the measured $c^\dagger$ and $c$ operators will retain a factor identical to $w^\text{box}$. 

In an inhomogeneous or symmetry-broken state, the effect of the driving field may be non-trivial, because the force 
may depend on the position or on the sub-lattice.

\subsection{Time-dependent Hubbard interaction}

We next consider an interaction quench, where $U(t)$ and $\mu(t)$ in Eq.~(\ref{H_loc}) are time-dependent, $\lambda$ and $\omega_0$ are constant, and $F=0$. 
In this case, the Lang-Firsov transformation defined by Eqs.~(\ref{LFtrafo}) and (\ref{x0_F}) becomes time-independent and the only time-dependence appears in $\tilde w_\text{Hubbard}$, since the shifted chemical potential and interaction parameters are $\tilde U(t)=U(t)-2\lambda^2/\omega_0$ and $\tilde \mu(t)=U(t)-\lambda^2/\omega_0$. To formulate the algorithm, we can simply set $F=F'=0$ in the equations of the previous subsection, which in particular means that the ``box" terms in the weight and in 
Fig.~\ref{illustration} disappear. 
The interaction quench calculation is therefore a straightforward generalization of the equilibrium algorithm\cite{Werner2007} to the Kadanoff-Baym contour.

\subsection{Time-dependent phonon coupling}

A more complicated situation arises if the electron-phonon coupling strength $\lambda$ is time-dependent: 
\begin{eqnarray}
H_\text{loc}(t) &=& U n_\uparrow n_\downarrow-\mu(n_\uparrow+n_\downarrow)\nonumber\\
&& +\lambda(t) (n_\uparrow+n_\downarrow-1)(b^\dagger+b)+\omega_0b^\dagger b.
\label{H_loc_lambda}
\end{eqnarray}
In this case, the standard Lang-Firsov transformation (\ref{LFtrafo}) simply replaces the electron-phonon coupling term $\lambda(t)X n$ by another one of the type $-\lambda'(t)P n$. We hence apply a generalized unitary transformation of the form
\begin{equation}
W(t)=e^{i(PX_0(t)+XP_0(t))},
\end{equation}
which implies, as usual,
\begin{align}
W^\dagger(t)X W(t) &= X-X_0(t), \\
W^\dagger(t)P W(t) &= P+P_0(t). 
\end{align}
The nontrivial relation is
\begin{align}
W^\dagger(t+\Delta t)W(t)=&e^{-iX(t) P'_0(t)\Delta t}e^{-iP(t) X'_0(t)\Delta t}\nonumber\\
&\times e^{\frac{i}{2}(P'_0(t)X_0(t)-X'_0(t)P_0(t))\Delta t}.
\end{align}
After the transformation, the terms $\frac{\omega_0}{2}(X^2+P^2)+\sqrt{2}\lambda(t)(n_\uparrow+n_\downarrow-1)X$ in Eq.~(\ref{H_loc_lambda}) plus the time-dependent basis change yield
\begin{align}
&\frac{\omega_0}{2}((X\!-\!X_0)^2+(P\!+\!P_0)^2)\!+\!\sqrt{2}\lambda(t)(n_\uparrow\!+\!n_\downarrow-1)(X\!-\!X_0)\nonumber\\
&+XP'_0+PX'_0-\frac{1}{2}(P'_0X_0-X'_0P_0)\nonumber\\
&=\frac{\omega_0}{2}(X^2+P^2)+\frac{\omega_0}{2}(X_0^2+P_0^2)\nonumber\\
&-\frac{1}{2}(P'_0X_0-X'_0P_0)-\sqrt{2}\lambda(n_\uparrow+n_\downarrow-1)X_0\nonumber\\
&+(-\omega_0X_0+P'_0+\sqrt{2}\lambda(n_\uparrow+n_\downarrow-1))X\phantom{\frac{1}{2}}\nonumber\\
&+(\omega_0P_0+X'_0)P.\phantom{\frac{1}{2}}\label{eq_shifted}
\end{align}
To eliminate the electron-phonon coupling, we have to set the last two terms to zero:
\begin{align}
X'_0(t)&=-\omega_0 P_0(t), \label{eqx}\\
P'_0(t)&=\omega_0 X_0(t)-f(t),\label{eqp}
\end{align}
where we have introduced the abbreviation 
\begin{align}
f(t)&=\sqrt{2}\lambda(t)(n_\uparrow+n_\downarrow-1).
\end{align}
First of all, we note that if $\lambda$ is time-independent, the solution of Eqs.~(\ref{eqx}), (\ref{eqp}) consistent with the initial condition is $P_0=0$, $X_0=\frac{1}{\omega_0}f=\sqrt{2}\frac{\lambda}{\omega_0}(n_\uparrow+n_\downarrow-1)$, in agreement with Eq.~(\ref{x0_F}). In the general case, where $\lambda$ is time-dependent, the solution becomes
\begin{align}
X_0(t)&=\frac{f(0)}{\omega_0}\cos(\omega_0t)+\int_0^td\bar t \sin(\omega_0(t-\bar t))f(\bar t),\label{xt}\\
P_0(t)&=\frac{f(0)}{\omega_0}\sin(\omega_0t)-\int_0^td\bar t \cos(\omega_0(t-\bar t))f(\bar t).\label{pt}
\end{align}
Plugging Eqs.~(\ref{eqx}), (\ref{eqp}), (\ref{xt}) and (\ref{pt}) into Eq.~(\ref{eq_shifted}), we find, besides a decoupled phonon term $\frac{\omega_0}{2}(X^2+P^2)$ an electronic term
\begin{align}
&\frac{\omega_0}{2}(X_0^2+P_0^2)+\frac{1}{2}(-\omega_0 X_0^2+f X_0-\omega_0P_0^2)-fX_0\nonumber\\
&=-\frac{1}{2}fX_0 \equiv (1+2n_\uparrow n_\downarrow-(n_\uparrow+n_\downarrow))g(t),\\
&g(t)=-\frac{\lambda(t)\lambda(0)}{\omega_0}\cos(\omega_0 t)-\lambda(t)\int_0^t d\bar t \lambda(\bar t)\sin (\omega_0 (t-\bar t)).
\end{align} 
This means that the interaction and chemical potential are shifted as 
\begin{align}
U&\rightarrow \tilde U(t)=U+2g(t),\\
\mu&\rightarrow \tilde U(t)=\mu+g(t).
\end{align}
For the shift at $t=0$, $g(0)=-\frac{\lambda(0)^2}{\omega_0}$, one recovers the well-known formulas for the transformed Holstein-Hubbard model 
with time-independent couplings, $\tilde U=U-\frac{2\lambda(0)^2}{\omega_0}$ and $\tilde \mu=\mu-\frac{\lambda(0)^2}{\omega_0}$.

Because the operator $n_\sigma$ is time-independent in the transformation, we have
\begin{align}
W(t)&=e^{i(PX_0(t)+XP_0(t))}\nonumber\\
&=e^{i(P\tilde X_0(t)+X\tilde P_0(t))\sqrt{2}(n_\uparrow +n_\downarrow-1)},\\
\tilde X_0(t)&=\frac{\lambda(0)}{\omega_0}\cos(\omega_0 t)+\int_0^t d\bar t \sin(\omega_0(t-\bar t))\lambda(\bar t),\\
\tilde P_0(t)&=\frac{\lambda(0)}{\omega_0}\sin(\omega_0 t)-\int_0^t d\bar t \cos(\omega_0(t-\bar t))\lambda(\bar t).
\end{align}
We use these expressions to find the transformation of the fermionic creation and annihilation operators:
\begin{align}
\tilde c_\sigma &= W^\dagger(t)c_\sigma W(t) = e^{i(P\tilde X_0(t)+X\tilde P_0(t))\sqrt{2}} c_\sigma \nonumber\\
&= e^{-((b^\dagger-b)\tilde X_0(t)+i(b^\dagger+b)\tilde P_0(t))}c_\sigma\nonumber\\
&\equiv e^{-(\gamma(t)b^\dagger-\gamma^*(t)b)}c_\sigma,
\end{align}
with 
\begin{align}
\gamma(t)
&=\tilde X_0(t)-i\tilde P_0(t)
\nonumber \\
&=\frac{\lambda(0)}{\omega_0}e^{-i\omega_0 t}+i\int_0^t d\bar t e^{-i\omega_0(t-\bar t)}\lambda(\bar t).
\end{align}
Similarly,
\begin{align}
\tilde c^\dagger_\sigma &= W^\dagger(t)c^\dagger_\sigma W(t) = e^{+(\gamma(t)b^\dagger-\gamma^*(t)b)}c^\dagger_\sigma.
\end{align}
After the separation of the electron and phonon contributions we must therefore evaluate a trace over a sequence of phonon-operators $O_b=e^{\pm(\gamma(t)b^\dagger-\gamma^*(t)b)}$ with the time-evolution between operators given by $H_b=\frac{\omega_0}{2}(X^2+P^2)=\omega_0 b^\dagger b$. In an interaction representation, this trace factor becomes $\text{Tr}[e^{-\beta H_b}T O_b(t_{2n})\ldots O_b(t_1)]$, where the operators in the interaction representation are
\begin{align}
&O_b(t)=e^{\pm(\gamma(t)b^\dagger(t)-\gamma^*(t)b(t))},\\
&b^\dagger(t)=e^{i\omega_0t}b^\dagger, \quad b(t)=e^{-i\omega_0t}b.
\end{align}
Splitting $\gamma(t)$ into a modulus and a phase, $\gamma(t)\equiv r(t)e^{i\phi(t)}$ we can write 
$w_b$ in a form analagous to Eq.~(\ref{wbF0}), with the substitutions
\begin{align}
&\tilde t_n=t_n+\frac{1}{\omega_0}\phi(t_n)\nonumber\\
&\tilde s_n=s_nr(t_n),
\end{align}
and $s_n=\pm 1$. After some straight-forward algebra, this leads to the expression
\begin{align}
w_b&=\exp\Bigg[\frac{-1}{\sinh(\frac{\beta\omega_0}{2})}\sum_{n>m}\frac{s_ns_m}{2}\Big(\gamma^*(t_n)\gamma(t_m)\nonumber\\
&\times e^{(\frac{\beta}{2}-i(t_n-t_m))\omega_0}+\gamma(t_n)\gamma^*(t_m)e^{-(\frac{\beta}{2}-i(t_n-t_m))\omega_0}\nonumber\\
&-(\gamma(t_n)\gamma^*(t_n)+\gamma(t_m)\gamma^*(t_m))\cosh\big(\frac{\beta\omega_0}{2}\big)
\Big)\Bigg].
\label{line_weight_lambda}
\end{align}
There are no ``box" terms in the bosonic weight, and we can directly read off the ``line" weights from Eq.~(\ref{line_weight_lambda}).

As a first example, 
we consider a quench of the phonon-coupling from $\lambda_1$ (at $t=0$) to $\lambda_2$ (at $t>0$). In this case, one finds $g(0)=-\frac{\lambda_1^2}{\omega_0}$, $\gamma(0)=\frac{\lambda_1}{\omega_0}$, and for $t>0$
\begin{align}
&g(t)=-\frac{\lambda_2^2}{\omega_0}-\frac{\lambda_2(\lambda_1-\lambda_2)}{\omega_0}\cos(\omega_0t),\\
&\gamma(t)=\frac{\lambda_1-\lambda_2}{\omega_0}e^{-i\omega_0 t}+\frac{\lambda_2}{\omega_0},
\end{align}
which means that $g(t)$ and hence $\tilde U(t)=U+2g(t)$ oscillate forever, except for $\lambda_2=0$. 
Note that this does not necessarily imply that the system will not relax, due to the effect of the nonlocal couplings in time ($w^\text{line}$).  

\begin{figure}[t]
\begin{center}
\includegraphics[angle=-90, width=0.49\columnwidth]{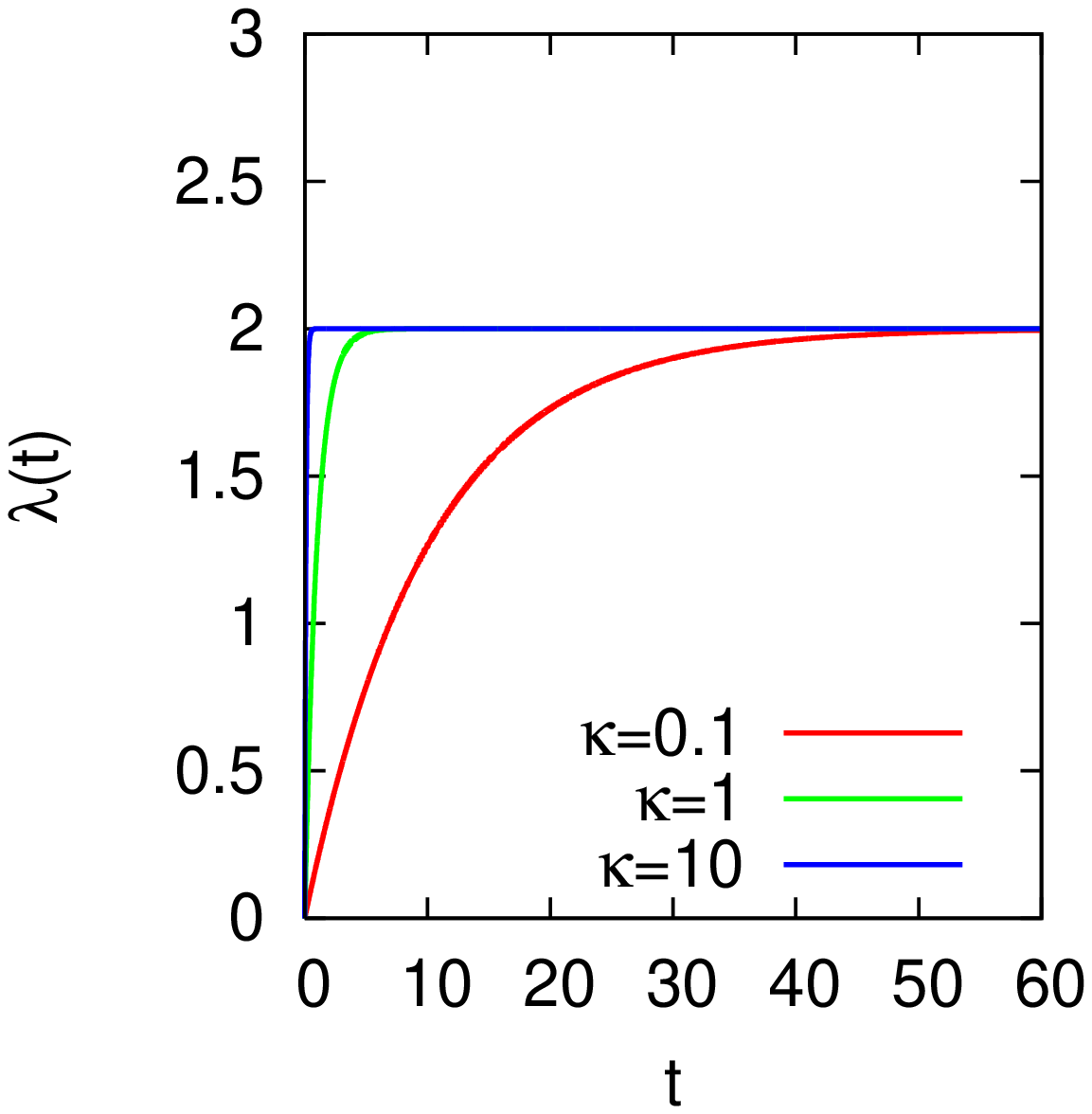}
\includegraphics[angle=-90, width=0.49\columnwidth]{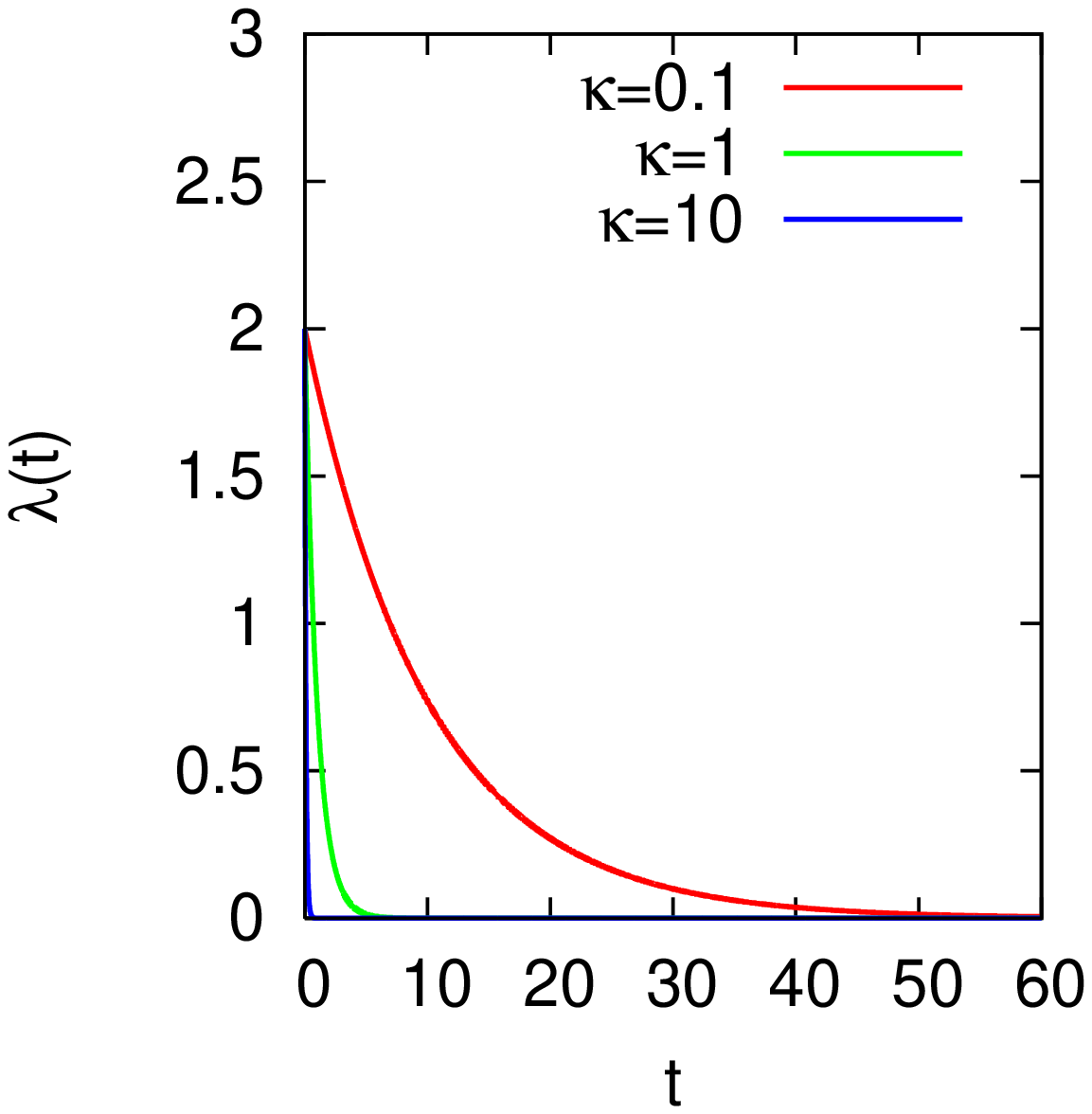}\\
\includegraphics[angle=-90, width=0.49\columnwidth]{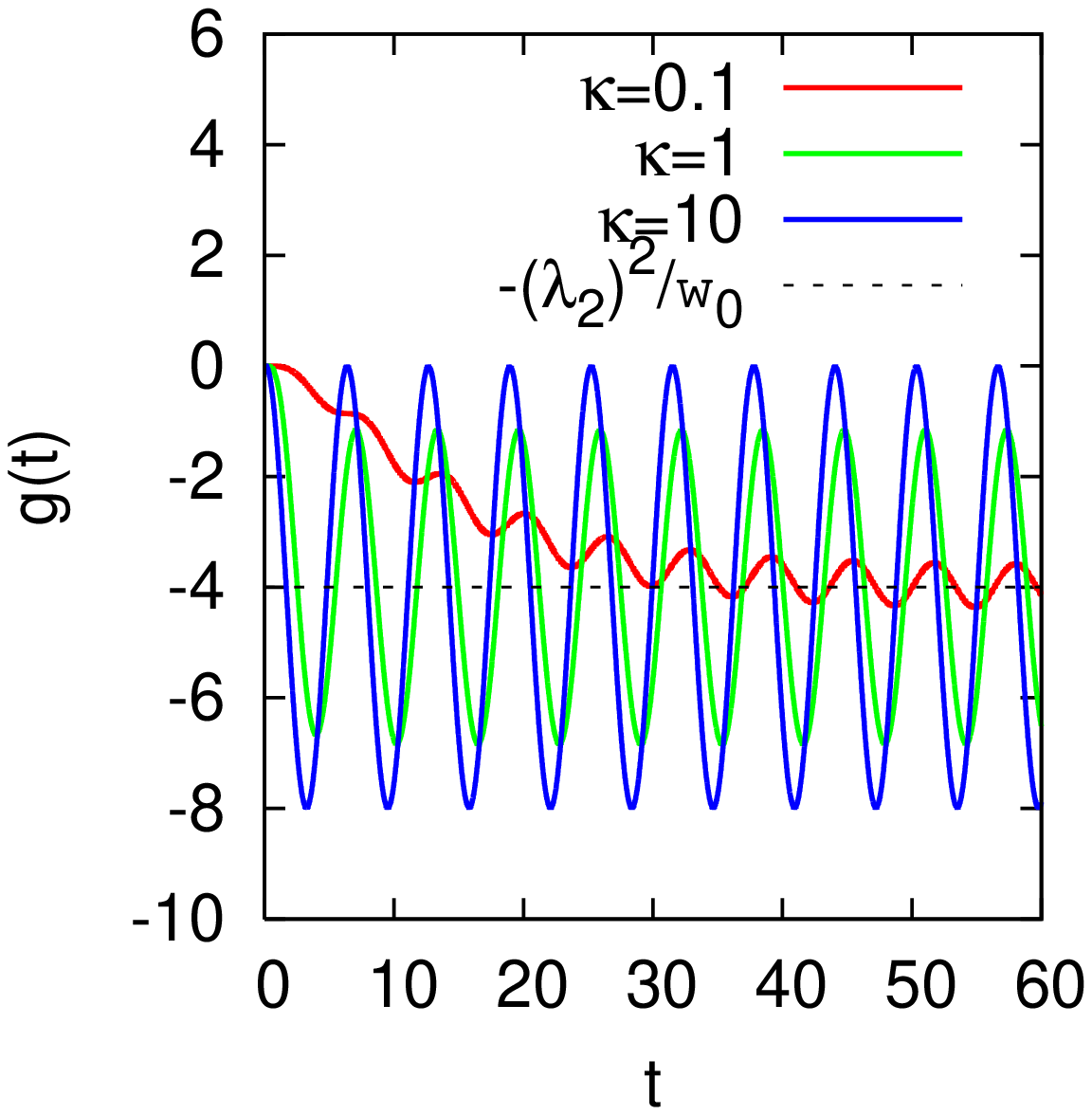}
\includegraphics[angle=-90, width=0.49\columnwidth]{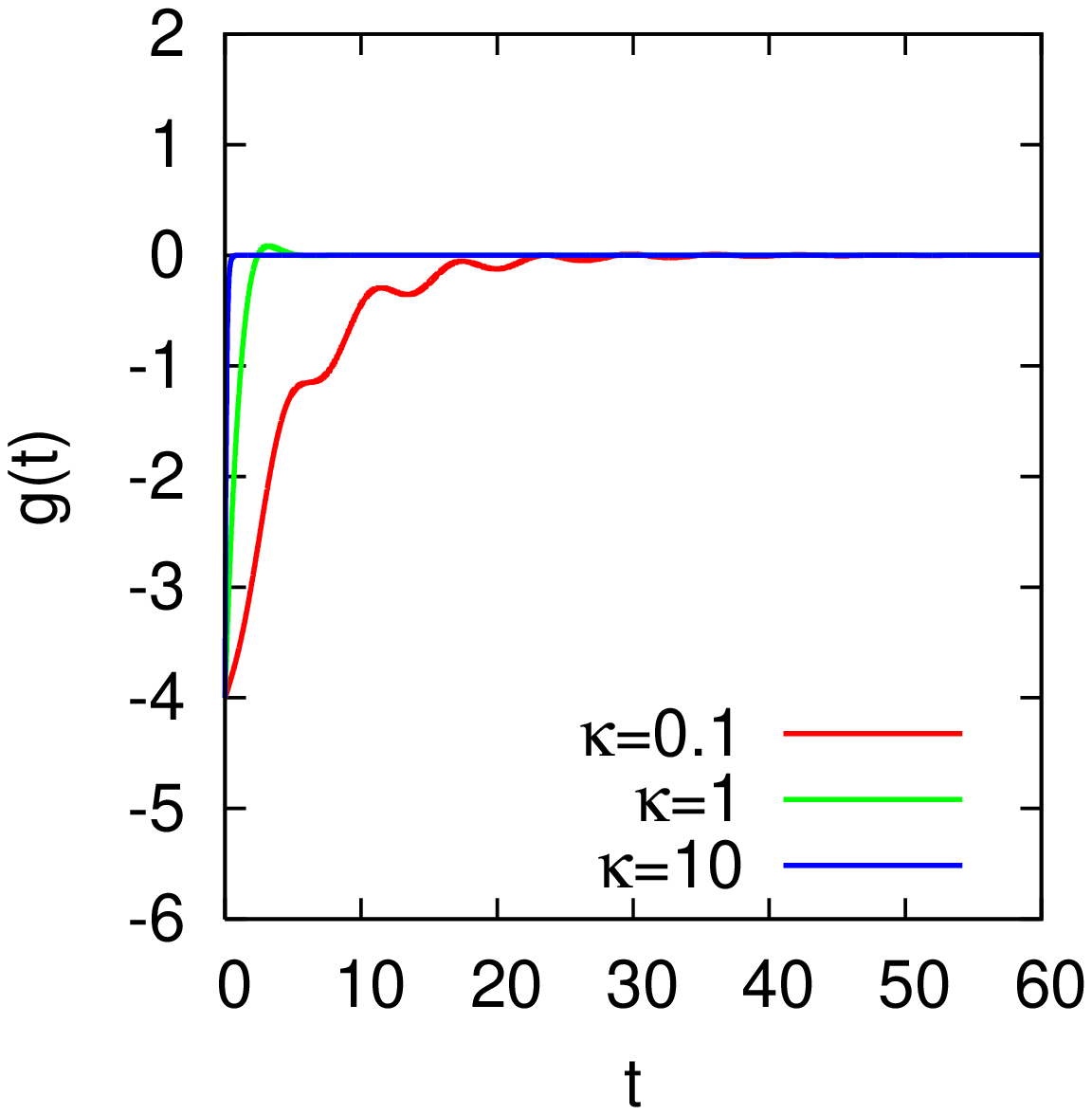}\\
\caption{Time-evolution of the phonon coupling $\lambda(t)$ and of the change in the effective instantaneous interaction $g(t)$ for the exponential switching (\ref{exp_relax}) and indicated values of $\kappa$ ($\omega_0=1$). Left panels: ramping the phonon coupling up from $0$ to $2$. Right panels: ramping the phonon coupling down from $2$ to $0$.}
\label{fig_g}
\end{center}
\end{figure}

In our calculations, we will consider an exponential switching from $\lambda_1$ to $\lambda_2$ on a time-scale controlled by the parameter $\kappa$:
\begin{equation}
\lambda(t)=\lambda_2+(\lambda_1-\lambda_2)e^{-\kappa t}.
\label{exp_relax}
\end{equation}
In this case, the behavior of $g(t)$ is qualitatively different in the regimes $\kappa\ll \omega_0$ and $\kappa \gg \omega_0$, as illustrated in Fig.~\ref{fig_g}. A fast switching to a nonzero $\lambda$ leads to large-amplitude oscillations in $g(t)$, similar to the case of the quench, while in a slow switching process, the oscillations are suppressed and the $g(t)$ approaches the value expected in the final equilibrium state ($-\lambda_2^2/\omega_0$) more or less smoothly (see Fig.~\ref{fig_g}). 
  
\subsection{Time-dependent phonon frequency}

Finally, let us discuss the situation where the phonon-coupling $\omega_0(t)$ is time-dependent, while $\lambda$ and $U$ are fixed and $F=0$. To decouple the electrons and phonons in this case, we may proceed as in the phonon-coupling quench section. Instead of Eqs.~(\ref{eqx}) and (\ref{eqp}) we obtain
\begin{align}
X'_0(t)&=-\omega_0(t) P_0(t), \label{eqxw0}\\
P'_0(t)&=\omega_0(t) X_0(t)-f,\label{eqpw0}
\end{align}
with $f=\sqrt{\lambda}(n_\uparrow+n_\downarrow-1)$ time-independent. By introducing the integral
\begin{equation}
w(t)=\int_0^t \omega_0(t')dt',
\end{equation}
we can write the solution of Eqs.~(\ref{eqxw0}) and (\ref{eqpw0}) as
\begin{align}
X_0(t)&=\frac{f}{\omega_0(0)}\cos(w(t))+\int_0^td\bar t \sin(w(t)-w(\bar t))f,\label{xtw0}\\
P_0(t)&=\frac{f}{\omega_0(0)}\sin(w(t))-\int_0^td\bar t \cos(w(t)-w(\bar t))f\label{ptw0},
\end{align}
which leads to the time-dependent shifted parameters
\begin{align}
U&\rightarrow \tilde U(t)=U+2g(t),\\
\mu&\rightarrow \tilde U(t)=\mu+g(t),\\
g(t)&=-\frac{\lambda^2}{\omega_0(0)}\cos(w(t))-\lambda^2\int_0^t d\bar t \sin (w(t)-w(\bar t)).
\end{align}
The fermionic and bosonic creation and annihilation operators transform as
\begin{align}
\tilde c_\sigma^\dagger &= e^{+(\gamma(t)b^\dagger-\gamma^*(t)b)}c^\dagger_\sigma,\\
\tilde c_\sigma &= e^{-(\gamma(t)b^\dagger-\gamma^*(t)b)}c_\sigma,\\
b^\dagger(t)&=e^{iw(t)}b^\dagger,\\
b(t)&=e^{-iw(t)}b,
\end{align}
with 
\begin{align}
\gamma(t)=\frac{\lambda}{\omega_0(0)}e^{-iw(t)}+i\lambda\int_0^t d\bar t e^{-i(w(t)-w(\bar t))}.
\end{align}
Splitting $\gamma(t)$ into a modulus and a phase, $\gamma(t)\equiv r(t)e^{i\phi(t)}$, and using the analogy to Eq.~(\ref{wbF0}) we find the weight
\begin{align}
w_b=&\exp\Bigg[\frac{-1}{\sinh(\frac{\beta\omega_0(0)}{2})}\sum_{n>m}\frac{s_ns_m}{2}\nonumber\\
&\times\Big(\gamma^*(t_n)\gamma(t_m)e^{(\frac{\beta}{2}\omega_0(0)-i(w(t_n)-w(t_m)))}\nonumber\\
&+\gamma(t_n)\gamma^*(t_m)e^{-(\frac{\beta}{2}\omega_0(0)-i(w(t_n)-w(t_m)))}\nonumber\\
&-(\gamma(t_n)\gamma^*(t_n)+\gamma(t_m)\gamma^*(t_m))\cosh\Big(\frac{\beta\omega_0(0)}{2}\Big)
\Big)\Bigg],
\label{line_weight}
\end{align}
from which one can read off the weight of a boson ``line" in the strong-coupling diagrams.

As a simple example, let us consider a quench from $\omega_0(t=0)=\omega_1$ to $\omega_0(t>0)=\omega_2$. In this case we find 
\begin{equation}
g(t)=-\frac{\lambda^2}{\omega_2}+\Big(\frac{\lambda^2}{\omega_2}-\frac{\lambda^2}{\omega_1}\Big)\cos(\omega_2 t),
\end{equation}
which for $\omega_2\ne \omega_1$ again leads to a persistent modulation of the shifted interaction and chemical potential, similar to the case of the phonon-coupling quench.

\subsection{Approximate solution of the impurity problem}

The stochastic sampling of all the diagrams of the type illustrated in Fig.~\ref{illustration} via some Monte Carlo procedure in principle allows to obtain a numerically exact solution of the nonequilibrium DMFT equations. However, since the weights are in general complex, such a simulation suffers from a phase problem which becomes more and more severe as one increases the length ($t$) of the Kadanoff-Baym contour.\cite{Werner2009} This limits the Monte Carlo approach to rather short times. In order to reach longer times, it is useful to consider approximate impurity solvers based on self-consistent strong-coupling expansions.\cite{Keiter1971, Pruschke1989} These solvers have been shown to give qualitatively correct solutions for the nonequilibrium dynamics of the Hubbard model in the strong correlation regime.\cite{Eckstein2010nca} Here, we adapt the strong-coupling perturbation theory to the Holstein-Hubbard model, where the strong-coupling diagrams -- in addition to fermionic creation and annihilation operators linked by hybridization lines -- contain phonon lines between all pairs of operators.

\begin{figure}[t]
\begin{center}
\includegraphics[angle=0, width=0.7\columnwidth]{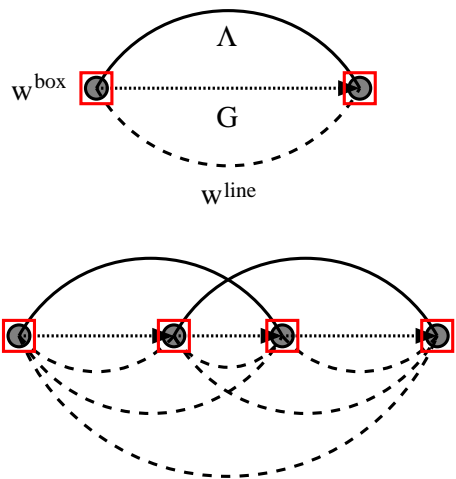}
\caption{Illustration of a diagram for the NCA pseudoparticle self-energy (top) and for the OCA pseudo-particle self-energy (bottom) in the simplest approximation.
Solid lines represent hybridization functions, dashed lines correspond to phonon mediated interactions, and the dotted arrows are boldified pseudoparticle Green's functions. The grey dots are either fermionic creation or annihilation operators (depending on the orientation of the hybridization lines).
}
\label{ncaoca}
\end{center}
\end{figure}

The simplest approximation is to multiply the hybridization function $\Lambda(t,t')$ by the weight of the phonon line: $\tilde \Lambda(t,t')=\Lambda(t,t')w^\text{line}(t,t')$, and to use this modified hybridization function in the lowest order strong-coupling perturbation theory (non-crossing approximation (NCA), see upper panel of Fig.~\ref{ncaoca} for an illustration of the pseudoparticle self-energy). In this approximation, only a relatively small number of phonon lines is retained. However, because of 
sign cancellations between the different phonon-line contributions (due to the factor $s_n s_m$ in Eqs.~(\ref{wbF0}), (\ref{line_weight_lambda}) and (\ref{line_weight}))
this approximation is less severe than it may seem. (For example, the weight of two phonon lines connecting a given operator to a distant creation/annihilation operator pair, whose separation on the contour is not too large, will almost cancel.)  While it is unclear if increasing the number of phonon lines at a given perturbation order in $\Lambda$ yields better results, 
one possible strategy would be 
to multiply the pseudo-particle propagators 
$g_\alpha$ (bare) and $G_\alpha$ (bold) by the weight of the phonon line: $\tilde g_\alpha(t,t')=g_\alpha(t,t')w^\text{line}(t,t')$, 
$\tilde G_\alpha(t,t')=G_\alpha(t,t')w^\text{line}(t,t')$, and to use these propagators within the usual NCA.

In the one-crossing approximation (OCA), the simplest scheme is represented by the self-energy sketched in the lower panel of Fig.~\ref{ncaoca}. Here, one could also capture more phonon lines by dressing the bare and bold pseudo-particle Green's function, as well as the hybridization lines and the self-energy itself, by phonon lines. 
Table~\ref{oca_table} compares the double occupancies obtained from the simplest NCA and OCA schemes to the exact Monte Carlo result. The results are for $U=10$, $\omega_0=0.2$, $1.0$ and increasing electron-phonon coupling strength $\lambda$ (in the Mott insulating phase, approaching the transition to the bipolaronic phase). 
As one can see, the OCA approximation reproduces the exact results rather well and correctly captures the interplay and competition between the electron-electron and electron-phonon interactions. The NCA approximation overestimates the interaction effects and leads to a shift of the phase boundary to the metallic phase. However, it still provides a qualitatively correct description of the strongly correlated (insulating) phases and we will thus use this particularly efficient approximation to investigate the real-time dynamics. We note that in these nonequilibrium studies, we are not interested in the very low-temperature/low-energy properties of the model, 
and in the applications below the initial states are thus rather accurately described within the NCA.

\begin{table}
\begin{tabular}{llll}
$\lambda$ & QMC & simple OCA & simple NCA \\
\hline
0 \hspace{1cm}    & 0.0051711 \hspace{1cm}  & 0.0051438 \hspace{1cm}\mbox{}&  0.0050026\\ 
0.2  & 0.0052004   & 0.0051599 & 0.0050145 \\
0.4  & 0.0052899   & 0.0052165 & 0.0050582 \\
0.6  & 0.0054545   & 0.0053333 & 0.0051527 \\
0.8  & 0.0058096   & 0.0056223 & 0.0054071 \\
0.9  & 0.010767        & 0.010476 & 0.010236 \\
0.95 & 0.047731    & 0.047292 & 0.047096\\
0.975 & 0.11869      & 0.11831 & 0.11827\\
1    & 0.24982     & 0.24982 & 0.25000 \\
1.1 & 0.49446     & 0.49451 & 0.49459\\
\\
0     & 0.0051715		      & 0.0051585                    & 0.0049969                   \\
0.4   & 0.0052190		      & 0.0051944                    & 0.0050320                   \\ 
0.8   & 0.0053715 		      & 0.0053112                    & 0.0051453                   \\
1.2   & 0.0056623		      & 0.0055381                    & 0.0053634                   \\
1.6   & 0.0061809		      & 0.0059470                    & 0.0057525                   \\
2   &	0.011291	      & 0.010664                    & 0.010401                   \\
2.1  & 0.034717		      & 0.033681                    & 0.033400                   \\
2.15  & 0.075718		      & 0.074567                    & 0.074379                   \\
2.2  & 0.16077		      & 0.16013                    & 0.16021                   \\
2.25  & 0.29053		      & 0.28643                    & 0.28673                   \\
\end{tabular}
\caption{
Comparison of the double occupancy for $\beta=5$, $U=10$ and $\omega_0=0.2$ (top), $\omega_0=1$ (bottom). To a good approximation, the phase transition from the Mott insulator to the bipolaronic insulator occurs when $\tilde U=U-2\lambda^2/\omega_0$ changes sign, and hence where the double occupancy crosses the noninteracting value of $0.25$. 
}
\label{oca_table}
\end{table}

\section{Results}
\label{results}

We will now illustrate the strong-coupling DMFT formalism and the simple NCA impurity solver with calculations of the time-evolution of the double occupancy and spectral function after a rapid parameter change. As we have mentioned, the application of external forces in a homogeneous system has no effect on local observables, such as the double occupancy, and only a 
trivial 
effect on nonlocal quantities (the DMFT result obtained in the absence of a force is multiplied by a force-dependent phase factor). We have also seen that the formalisms for
time-dependent phonon-coupling and time-dependent phonon-frequency
are very similar. Hence, we will concentrate here on two set-ups: (i) a $U$-pulse, which 
produces doublons and holes (similar to a photo-doping experiment)  
and allows us to study the {\it relaxation} of doublons in the presence of an electron-phonon coupling, and (ii) a $\lambda$-quench, which induces 
coherent phonon oscillations  
and allows us to investigate the phonon-enhanced {\it production} of doublons (similar to the case of ``modulation spectroscopy"\cite{Kollath2006}). 

The calculations are done for a semi-circular density of states of bandwidth $4v$ (self-consistency equation (\ref{bethe})) and we use $v$ [$v^{-1}$] as the unit of energy [time]. The phonon frequency will be fixed at $\omega_0=1$, which depending on the class of materials may seem rather high, but we are interested here only in qualitative aspects of electron-phonon coupled systems. We furthermore restrict our attention to the symmetric phases of the model.

\subsection{$U$-pulse: phonon enhanced doublon relaxation}

As a first application, we study the effect of phonons on the relaxation of artifically created doublons. In the Hubbard model, it is known\cite{Sensarma2010} that the relaxation time in the Mott insulating phase depends exponentially on the interaction $U$, and this dependence is clearly seen in DMFT calculations based on NCA or OCA solvers.\cite{Eckstein2010pump} The reason for the exponentially long doublon life-time in the strong-correlation regime is that the doublon-hole recombination releases an energy of order $U$, which in the limit where $U$  is much larger than the kinetic energy can only be absorbed by high-order scattering processes. In the presence of phonons, there are additional relaxation channels, which involve a transfer of multiples of the phonon energy $\omega_0$ from the electronic system to the lattice. Here, we investigate how this affects the doublon life-time. 

\begin{figure}[t]
\begin{center}
\includegraphics[angle=-90, width=0.9\columnwidth]{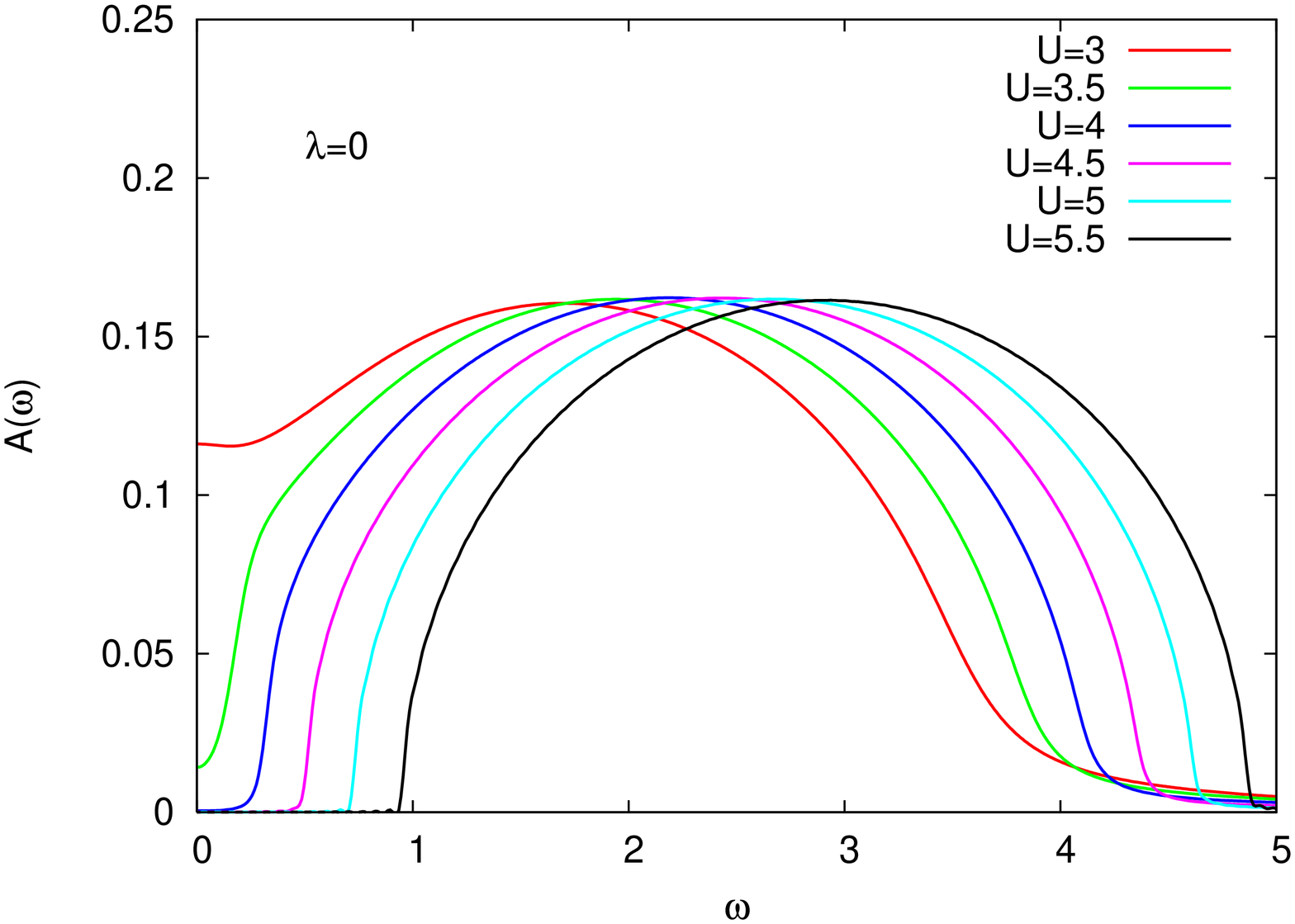}
\includegraphics[angle=-90, width=0.9\columnwidth]{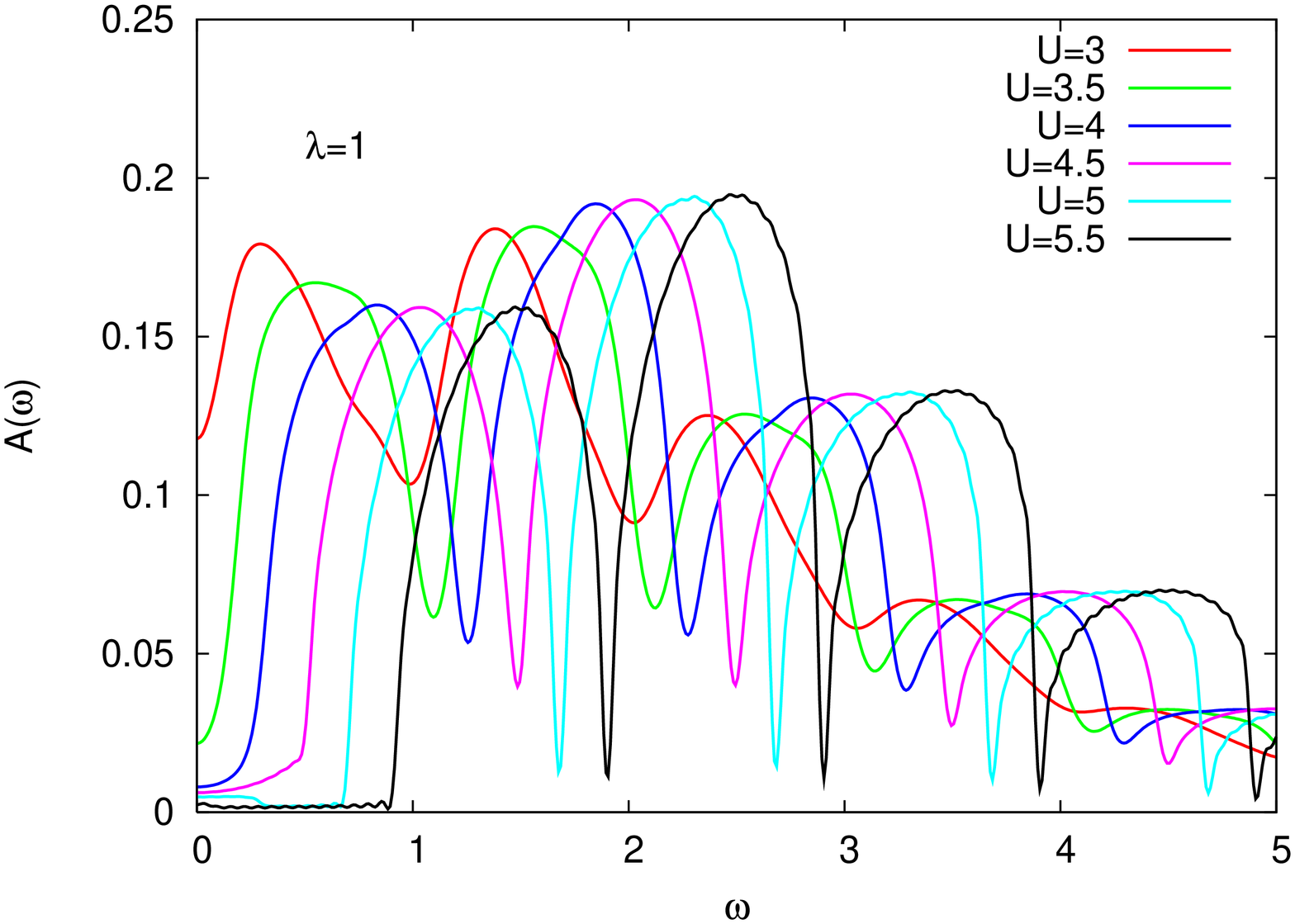}
\caption{Equilibrium spectral functions for $\lambda=0$ (top) and $\lambda=1$ (bottom) for indicated values of $U$. The inverse temperature is $\beta=5$.}
\label{fig_spectra}
\end{center}
\end{figure}   

To orient ourselves, we plot in Fig.~\ref{fig_spectra} the equilibrium spectral functions for inverse temperature $\beta=5$, $\lambda=0$ and $1$ and different values of $U$. In both the model with and without coupling to phonons, a gap opens around $U\approx 3.25$. The Hubbard bands in the paramagnetic calculation without phonons are relatively featureless and approach a semi-circle deep in the Mott insulating phase. (In the antiferromagnetic phase, the spectra would feature spin-polaron peaks.\cite{Taranto2012, Werner2012}) The phonon coupling leads to the formation of phonon-peaks with an energy separation of $\omega_0$. The gap size in the calculation with $\lambda=0$ is very similar to that for $\lambda=1$, which is a coincidence. However, in the model with phonon coupling, some spectral weight remains at the Fermi energy, even at $U=5$, because of the overlapping phonon side-bands. 

For the analysis of the data, it will be useful to define the gap size in the Holstein-Hubbard spectrum by the peak-to-peak separation between the first prominent side-bands (measured at the maxima). In this case one finds (for $\lambda=1$) that the gap is approximately $\omega_0$ for $U\approx 3.5$, $2\omega_0$ for $U\approx 4.5$ and $3\omega_0$ for $U\approx 5.5$. 

\begin{figure}[t]
\begin{center}
\includegraphics[angle=-90, width=0.9\columnwidth]{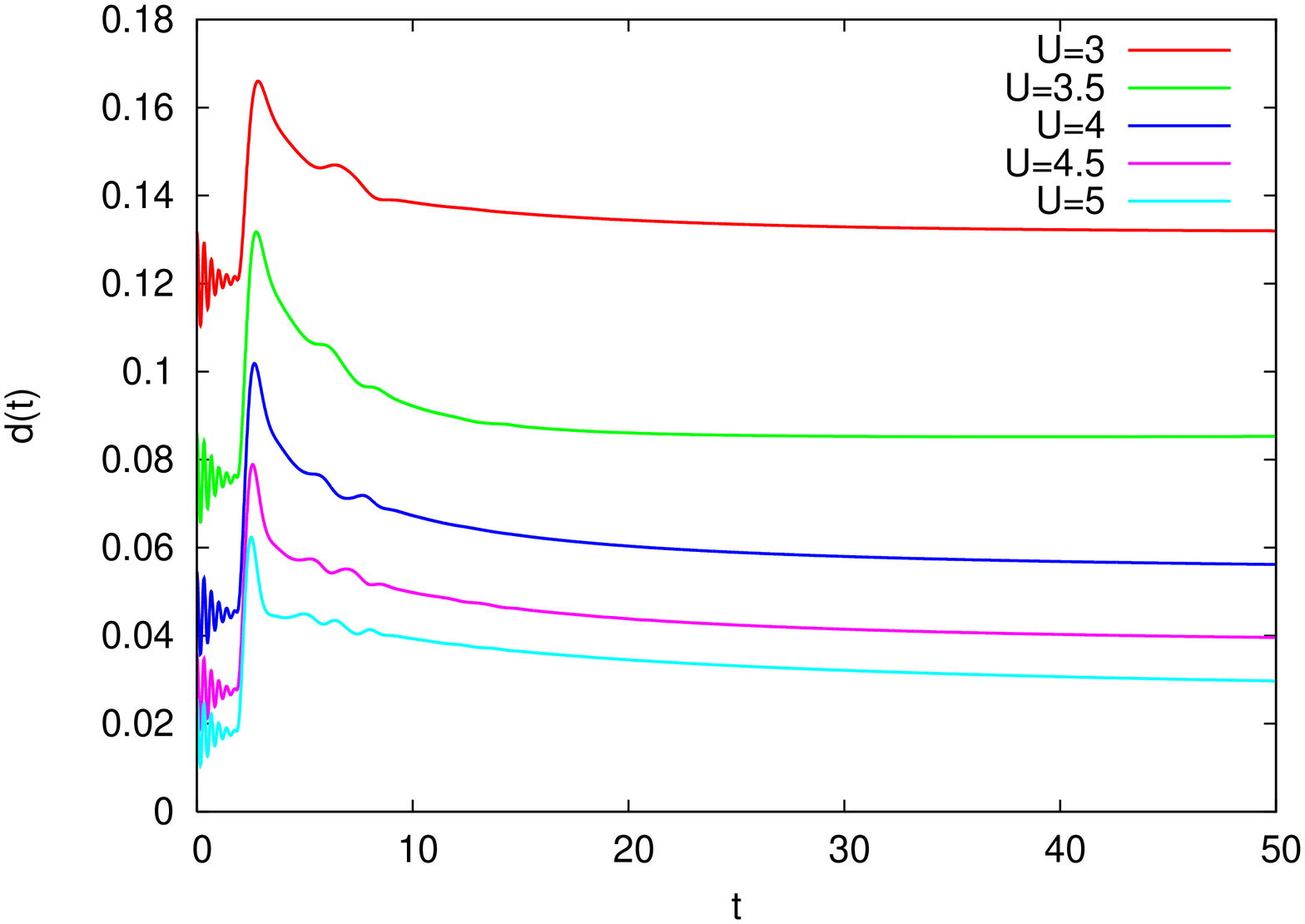}\\
\includegraphics[angle=0, width=0.9\columnwidth]{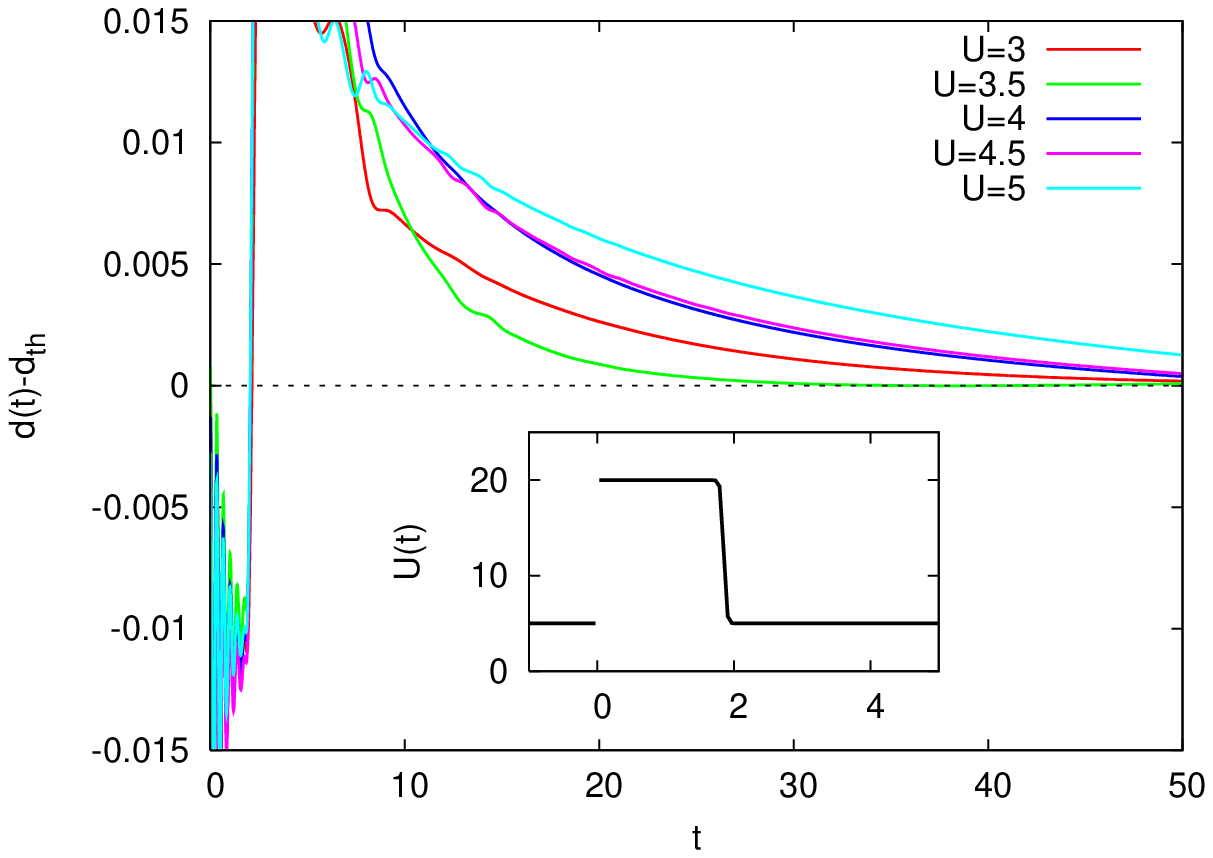}
\caption{Top panel: Time-evolution of the double occupancy after an interaction pulse of duration $t_p=2$ for different values of $U$ (initial temperature $\beta=5$, $\lambda=1$). Bottom panel: same data with $d_\text{th}$ (the value reached in the long-time limit) subtracted. The inset shows the form of the $U$-pulse for $U=5$.}
\label{fig_double}
\end{center}
\end{figure}   

\begin{figure}[t]
\begin{center}
\includegraphics[angle=-90, width=0.9\columnwidth]{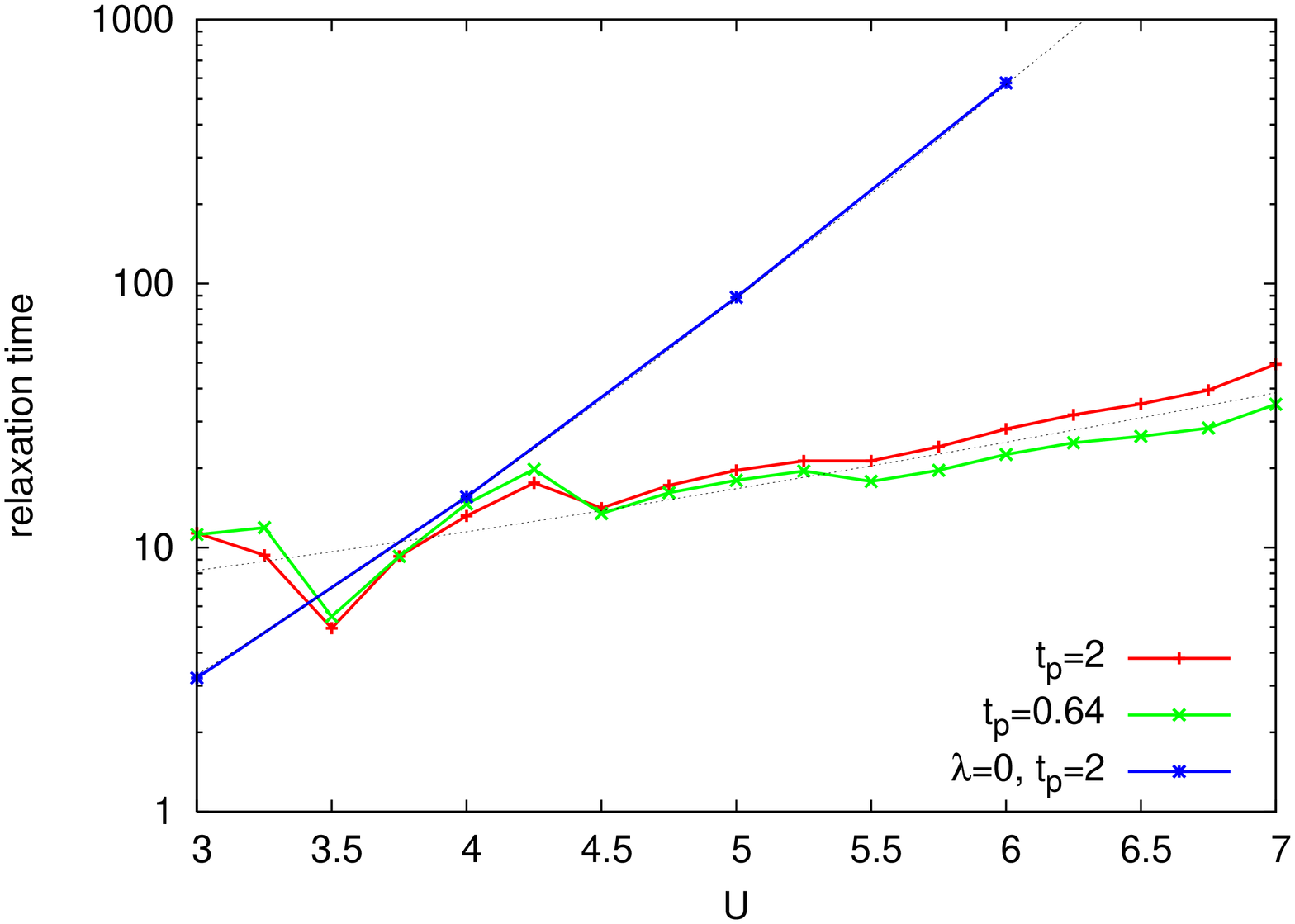}
\includegraphics[angle=-90, width=0.9\columnwidth]{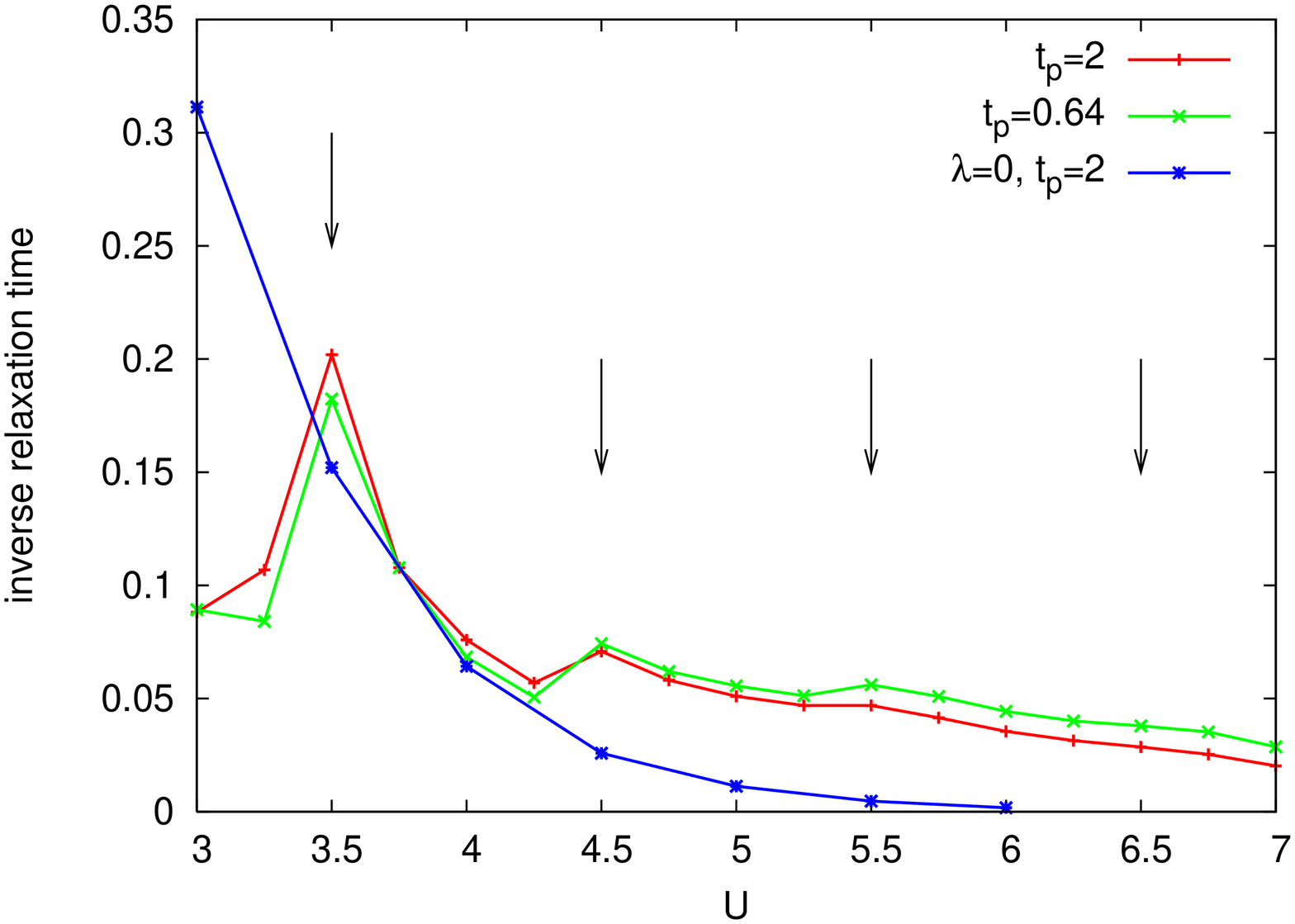}
\caption{Relaxation time (top) and inverse relaxation time (bottom) of the double occupancy after an interaction pulse of duration $t_p$ for different values of $U$. The results for $\lambda=1$ are compared to those obtained in the Hubbard model ($\lambda=0$, blue line). The initial temperature is $\beta=5$.
The values of $U$ for which the gap size in the equilibrium and time dependent spectral function becomes a multiple of $\omega_0$ are indicated by arrows.
}
\label{fig_tau}
\end{center}
\end{figure}   

In order to excite doublons, we apply an interaction pulse to the model with $\lambda=1$: the interaction $U(t)$ jumps from $U(0)=U$ to $U(0_+)=U_p=20$, and rapidly switches back to $U$ at $t=t_p$ (see inset in the lower panel of Fig.~\ref{fig_double}). We use a smooth switching at $t=t_p$ in order to improve the stability of the time-propagation scheme,\cite{Eckstein2010quenchlong} so that we can use a time-step $\Delta t=0.01$. The time-evolution of the double occupancy after a pulse of duration $t_p=2$ is shown in Fig.~\ref{fig_double}. The first quench to $U_p=20$ suppresses the double-occupancy and leads to rapid $1/U_p$ oscillations. After switching back to the initial $U$ at time $t\approx t_p$, the double occupancy shoots up to a value which is substantially larger than in the initial state (pulse-induced doublon-holon production) and then starts to relax towards the new thermal value. We fit the relaxation by an exponential function $d(t)=d_\text{th}+a\exp(-t/\tau)$ in the range $15\le t\le 50$ to extract the long-time limit $d_\text{th}$ and the relaxation time $\tau$. Close inspection of the top panel shows that the relaxation time is not a simple increasing function of $U$, as it is the case in the Hubbard model. This becomes even more evident when we subtract the fitted long-time value $d_\text{th}$ (lower panel). Here on can see that the relaxation for $U=3.5$ is substantially faster than for $U=3$, while the relaxation times for $U=4$ and $4.5$ is similar. Figure~\ref{fig_tau} plots the relaxation time $\tau$ and its inverse as a function of $U$, and clearly shows that the doublons relax fast for $U=3.5$, $4.5$, $5.5$, \ldots, i.e. whenever the gap-size in the equilibrium spectrum is a multiple of the phonon frequency. In this case, the emission of phonons provides an efficient relaxation pathway. 

A comparison to the relaxation time in the Hubbard model shows that the coupling to phonons leads to a substantially faster relaxation of doublons and to a much slower increase of the relaxation time with $U$, at least for interactions up to $U\approx 7$. The doublon relaxation time in the Hubbard model can be well fitted with the expected form\cite{Sensarma2010} $\tau=A\exp[\alpha U\log U]$ with $\alpha_{\lambda=0}=0.69$ (dashed line in the upper panel of Fig.~\ref{fig_tau}). If we apply the same fit to the Holstein-Hubbard case, $\alpha$ is reduced to $\alpha_{\lambda=1}\approx 0.15$, but 
it is obvious that the fit does not reproduce the resonance phenomena in the regime where the gap size is comparable to the phonon frequency. The much slower increase of the relaxation time with $U$ is due to the larger phase space for relaxation processes (combination of higher-order scattering processes and phonon emissions) and to the existence of a small density of states at the Fermi energy. At even larger $U$, when this density of states becomes exponentially small, and the doublon-holon recombination energy is much larger than the phonon energy, we do expect, as in the Hubbard case, an exponential dependence of $\tau$ on $U$, as described in Ref.~\onlinecite{Sensarma2010}. However, for $U>7$ it becomes difficult to measure a relaxation time, because the slow decrease of $d(t)$ is almost linear up to the longest accessible times. 

Figure~\ref{fig_tau} shows data for two different pulse lengths ($t_p=0.64$ and $2$), and thus for different excitation densities. For example, for $U=5$, the double occupancy in the intial state is $0.0256$, while the maximum double occupancy after the perturbation is $0.0813$ for $t_p=0.64$ and $0.0623$ for $t_p=2$ (a similar ratio between excitation densities is found at other values of $U$). Apparently, the relaxation time is not strongly dependent on the number of doublon-holon pairs produced by the excitation, and the observed resonance phenomena are independent of the excitation density.

\begin{figure}[t]
\begin{center}
\includegraphics[angle=-90, width=0.49\textwidth]{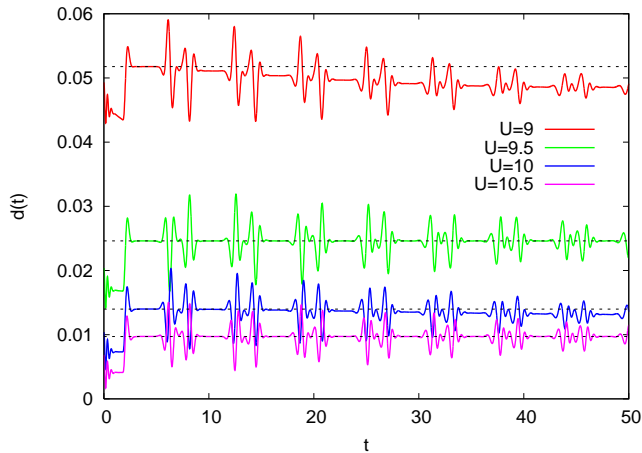}
\caption{Time-evolution of the double occupancy after a $U$-pulse of $t_p=2$ in the Mott insulator with $\lambda=2$ and indicated values of $U$. Dashed lines are plotted at the value corresponding to the first plateau after the pulse.
}
\label{fig_large_lambda}
\end{center}
\end{figure}

Finally, we show in Fig.~\ref{fig_large_lambda} the time-evolution of the double occupancy after a $U$-pulse of $t_p=2$ in the Mott insulator with strong electron-phonon coupling $\lambda=2$ and different values of $U$. In this case, the transition to the bipolaronic insulator occurs near $U=8$ and the separation between the first prominent phonon side-peaks is approximately $\omega_0$ for $U=9$, and $2\omega_0$ for $U=10$. Indeed we see that a relaxation of the double occupancy occurs on the time-scale of the plot for these two values of $U$, while no relaxation is evident for $U=9.5$ and $U=10.5$. A comparison with Fig.~\ref{fig_tau} however shows that the relaxation is much slower than in the more weakly correlated case with $\lambda=1$. For $U=9.5$, the double occupancy is essentially stuck even though the effective interaction ($\tilde U=1.5$) is small. Despite the strong screening of the interaction in equilibrium, the large ``bare $U$" seems to prevent a rapid relaxation of the doublons. Instead of a relaxation, the double occupancy exhibits ``echos" of the $U$-pulse perturbation, which are separated in time by one phonon oscillation period $2\pi\omega_0=6.28$. In the interaction regime where a phonon-enhanced relaxation is possible, the double-occupancy changes in a step-like manner after each echo event and it is not really possible to extract a relaxation time.

\subsection{$\lambda$-quench: phonon enhanced doublon production}

As a second example, we study the time-evolution after a rapid change in the electron-phonon coupling $\lambda$. We will consider the exponential switching (\ref{exp_relax}) from $\lambda_1=0$ to $\lambda_2>0$ with time-constant $\kappa=1$. This is a fast switching (``quench") in the sense that it leads to large-amplitude oscillations of $g(t)$ around $-2\lambda_2^2/\omega_0$, and thus to a strong periodic modulation of the effective interaction $\tilde U$ with frequency $\omega_0$.

\begin{figure}[t]
\begin{center}
\includegraphics[angle=-90, width=0.49\textwidth]{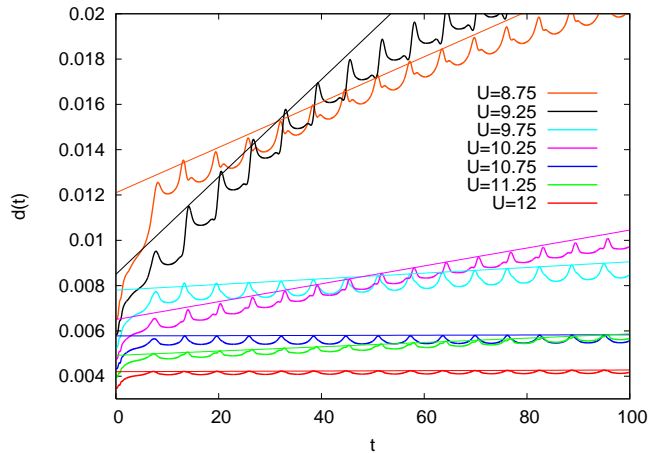}
\caption{Evolution of the double occupancy after a switch from $\lambda=0$ to $2$ ($\kappa=1$) for different values of $U$. The initial temperature is $\beta=5$. The straight lines are guides to the eye.
}
\label{fig_double_exp}
\end{center}
\end{figure}

The time-evolution of the double occupancy after a quench to $\lambda_2=2$ is shown in Fig.~\ref{fig_double_exp} for several values of $U$. In equilibrium, the transition to the bipolaronic insulator occurs near $\tilde U=U-2\lambda_2^2/\omega_0=0$, i.e. $U\approx 8$. Hence, the curves plotted in Fig.~\ref{fig_double_exp} are for quenches within the Mott insulating phase, but the smallest $U$ value is getting close to the bipolaronic phase boundary.

In the Hubbard model, an interaction quench from $U=12$ to $4$ would lead to a relatively fast exponential relaxation of the double-occupancy (see blue curve in Fig.~\ref{fig_tau} for the relaxation time in the Hubbard model after a different type of perturbation). However, in the Holstein-Hubbard case, where $\tilde U=4$ means that the very strong instantaneous repulsion $U=12$ is to a large extent compensated by a strong phonon-induced attraction $-2\lambda^2/\omega_0=-8$, the relaxation of the double occupancy towards the higher thermal value is seen to be very slow, which is a clear indication that these systems are more strongly correlated than a static description with interaction $\tilde U$ would suggest. This observation is consistent with the finding of Ref.~\onlinecite{Casula2012}, which showed that in equilibrium and at low enough temperature, the proper static description for the Holstein-Hubbard model involves the interaction $\tilde U$ {\it and} a reduced bandwidth. For $\lambda=2$ the bandwidth reduction factor is $\exp(-\lambda^2/\omega_0)=0.02$. While the effective static description is not accurate in the present case of $\omega_0=1$, strong electron-phonon coupling, and strong excitation of the phonons, it nevertheless provides some insight into the observed slow dynamics. It is also important to note that after the quench, $\tilde U$ will oscillate between $12-2\cdot 6.8 =-1.6$ and $12-2\cdot 1.2=9.6$ (see Fig.~\ref{fig_g}, bottom left panel), so that the instantaneous interaction periodically switches from strongly repulsive to attractive. In such a situation, the interpretation of the dynamics in terms of an equilibrium model seems difficult. 
 
\begin{figure}[t]
\begin{center}
\includegraphics[angle=-90, width=0.9\columnwidth]{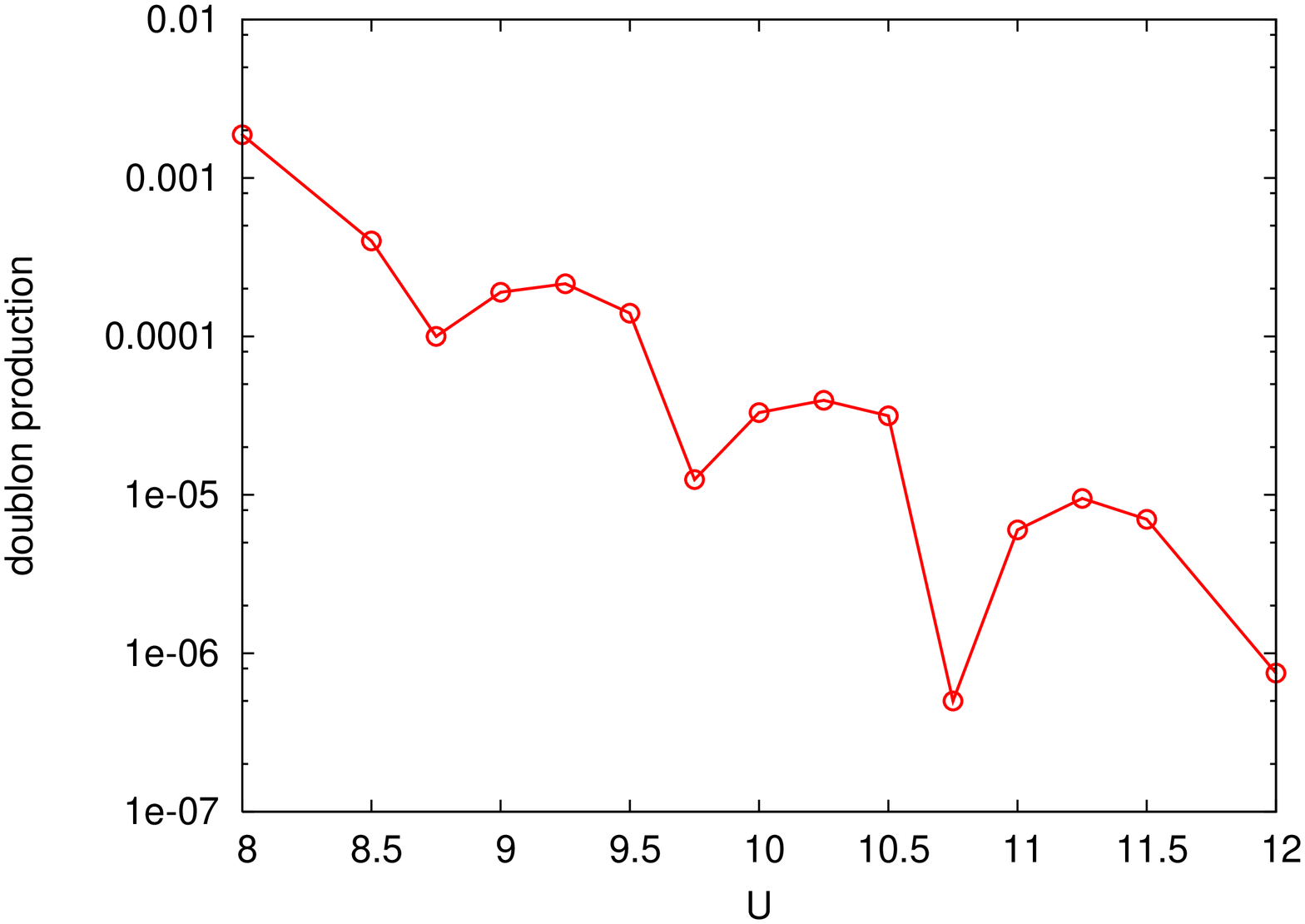}\\
\includegraphics[angle=-90, width=0.9\columnwidth]{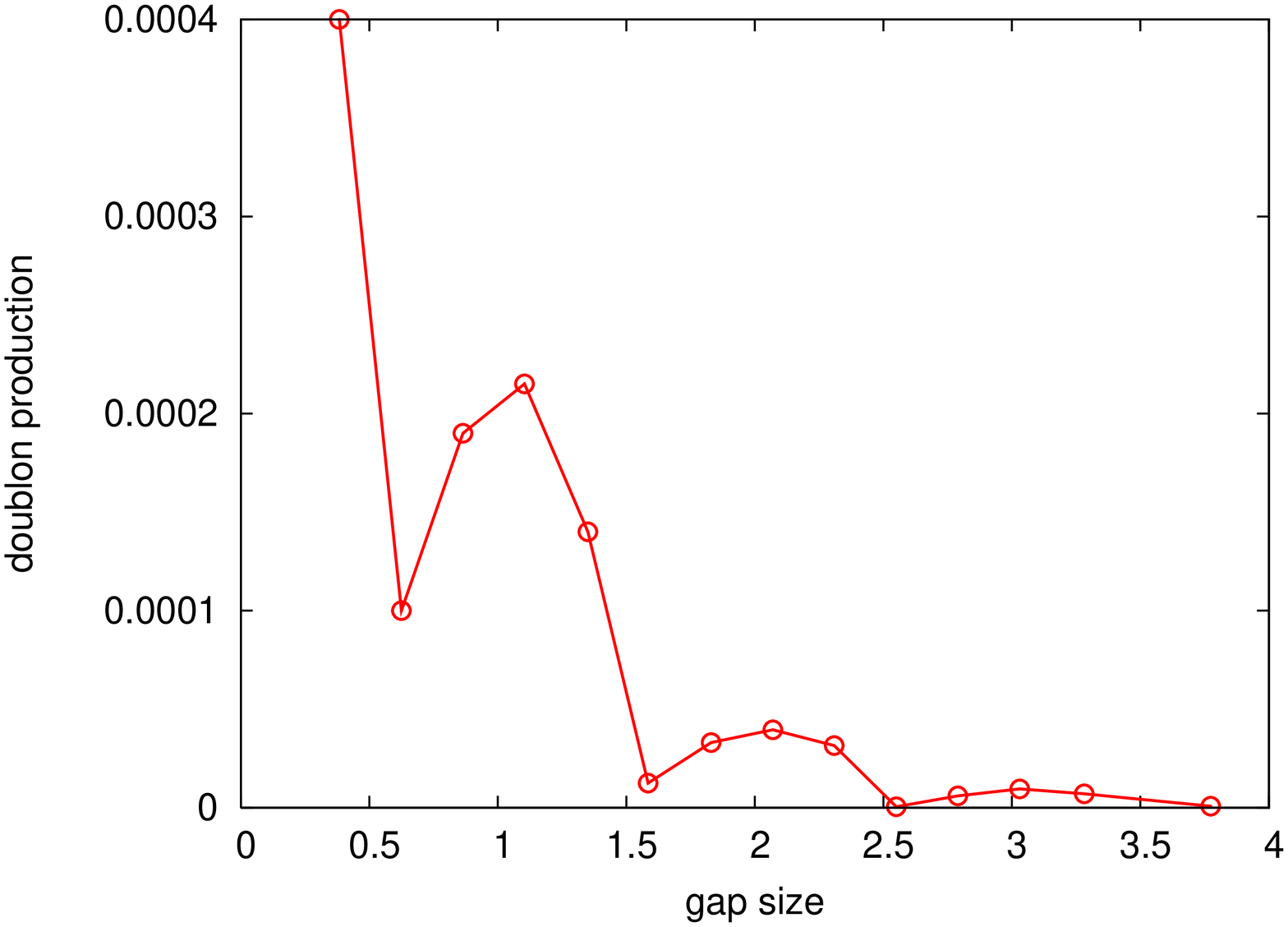}
\caption{
Top panel: Doublon production rate (slope of the roughly linear increase in $d(t)$) as a function of $U$ for quenches from $\lambda=0$ to $2$. The initial temperature is $\beta=5$.
Bottom panel: Doublon production rate as a function of the gap in the (nonequilibrium) spectral function.
}
\label{slope_exp}
\end{center}
\end{figure}     
   
It is also evident from Fig.~\ref{fig_double_exp} that the relaxation of the double occupancy, at least in the time interval which is plotted, is not exponential. Rather, the double occupancy increases roughly linearly, with superimposed oscillations that are almost undamped. These oscillations are due to the periodic modulation of the Lang-Firsov shifted interaction strength $\tilde U(t)$. A similar roughly linear increase in the double occupancy was found in nonequilibrium DMFT simulations of the Hubbard model with a periodically modulated $U$.\cite{Eckstein2010nca} The slopes of such curves are measured in modulation spectroscopy experiments on cold atom systems, to determine the Mott phase and the Mott gap (interaction strength $U$).\cite{Kollath2006,Joerdens2008} In the present case the coherent excitation of phonons by the $\lambda$-quench leads to a periodic oscillation of $\tilde U$ with frequency $\omega_0$, which in turn may enhance the production of doublons if the gap size is a multiple of the phonon frequency. 

As can be seen in Fig.~\ref{fig_double_exp}, the slope of a linear fit to the doublon curves (doublon production rate) exhibits a nontrivial dependence on $U$ and hence on the gap-size. For example, the production rate is substantially larger for $U=10.25$ than for $U=9.75$, even though the gap in the latter case is smaller. 
The slopes are plotted as a function of $U$ in the top panel of Fig.~\ref{slope_exp}. The doublon production is enhanced for $U\approx 9.25$, $10.25$ and $11.25$ and strongly suppressed for $U\approx 9.75$, $10.75$ and $11.75$. To understand this behavior and relate it to the gap size, we now analyze the spectral functions. 

\begin{figure}[t]
\begin{center}
\includegraphics[angle=-90, width=0.49\columnwidth]{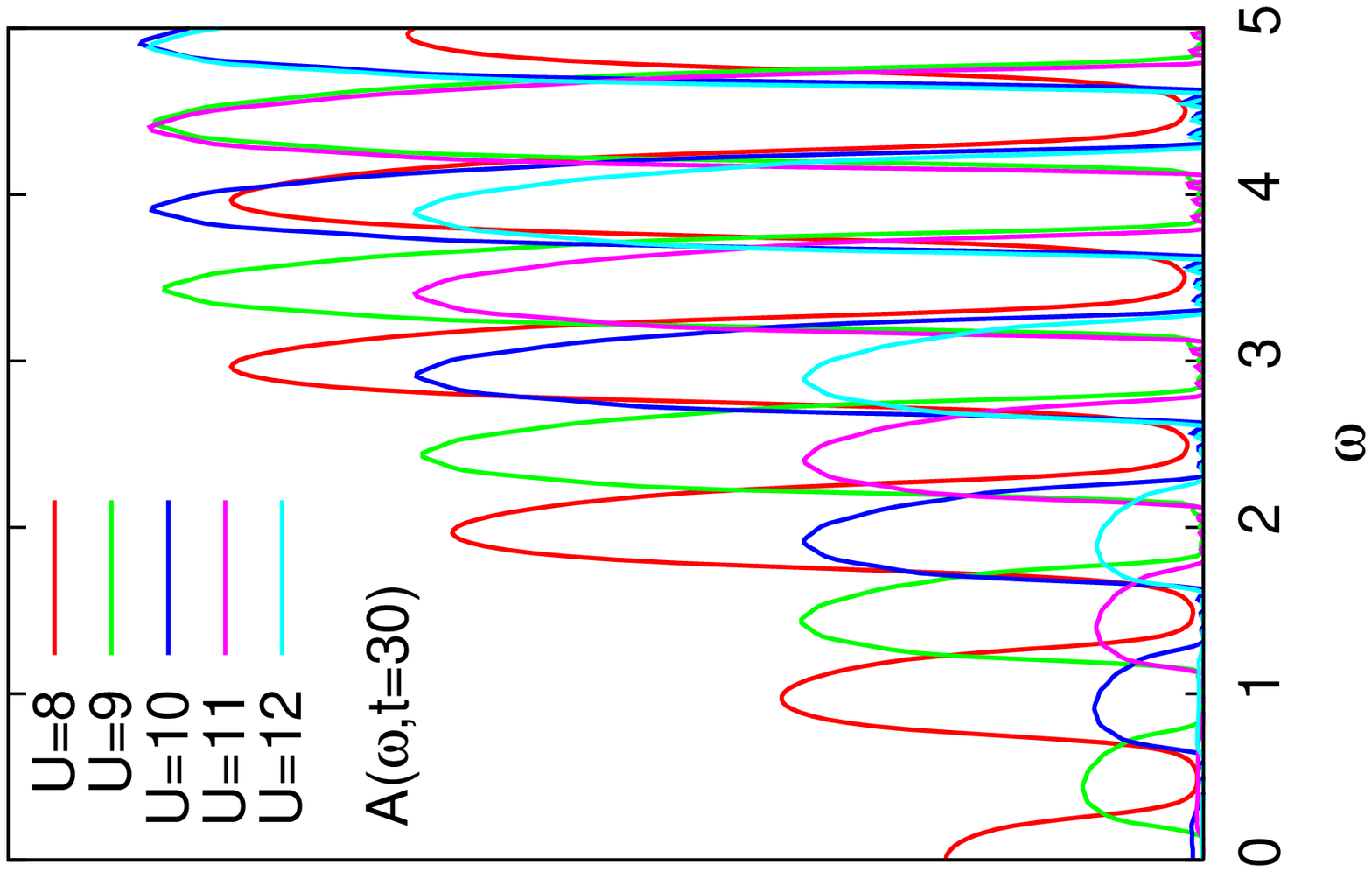}
\includegraphics[angle=-90, width=0.49\columnwidth]{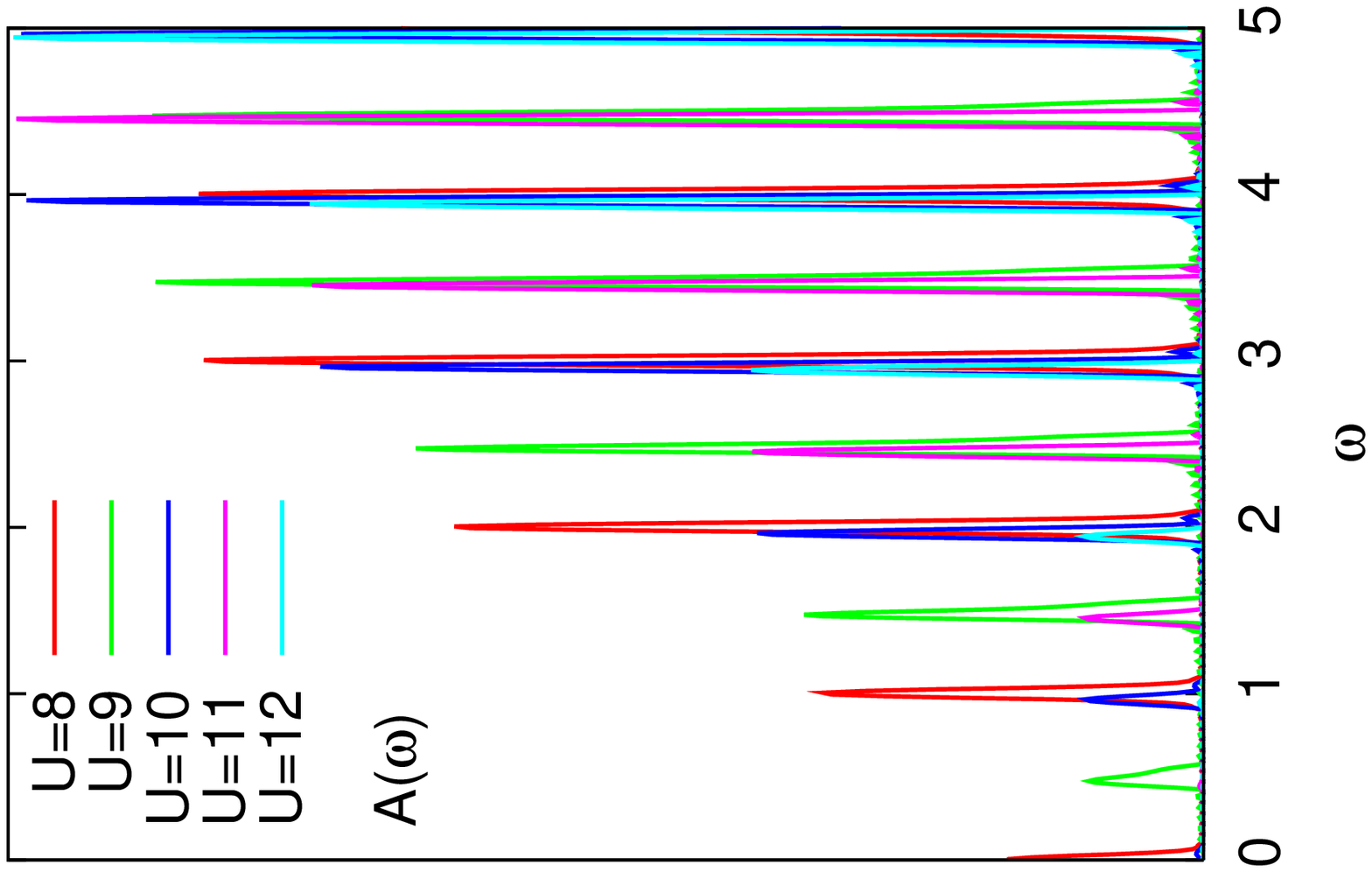}\\
\includegraphics[angle=-90, width=\columnwidth]{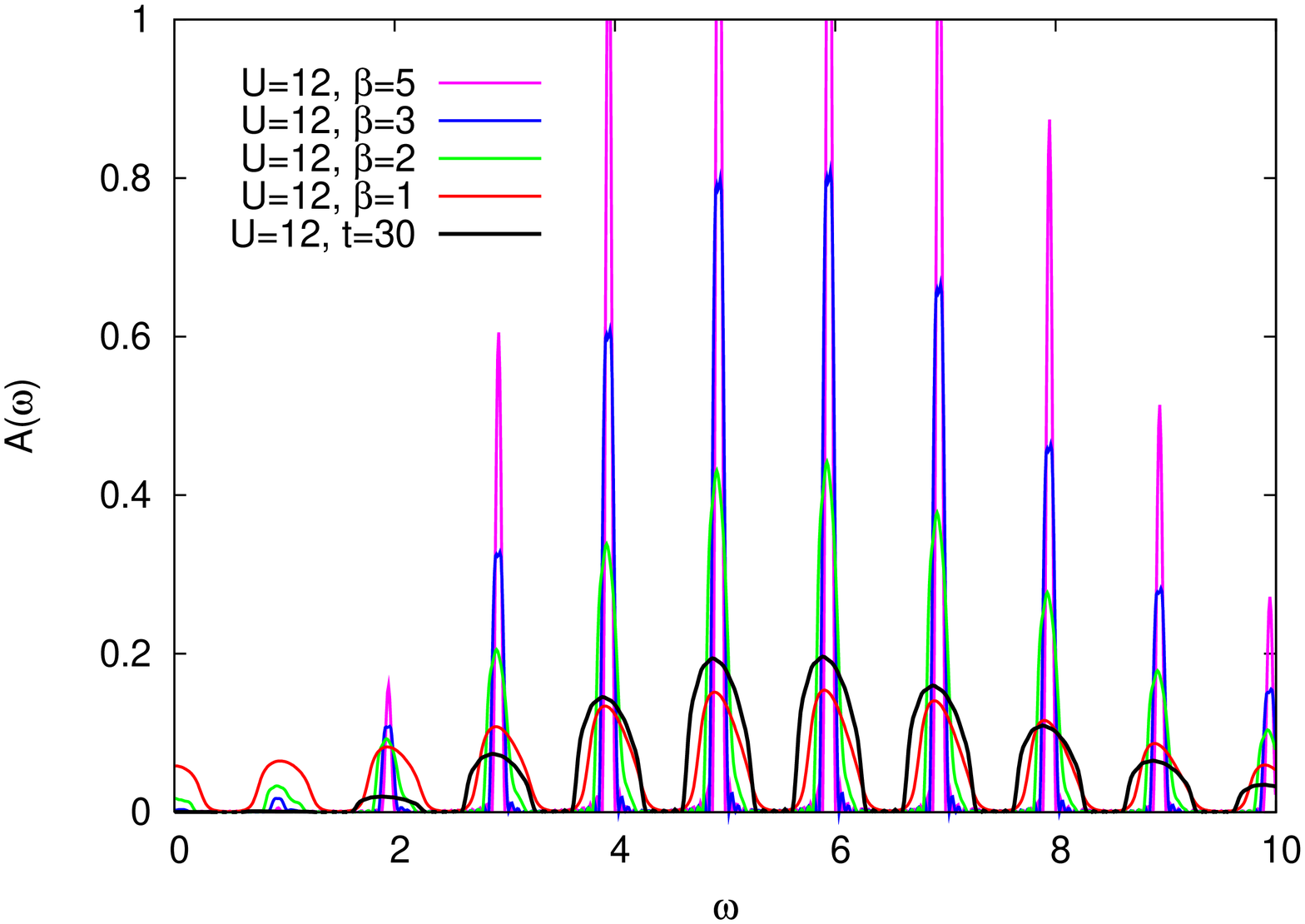}
\caption{
Top left panel: time-averaged spectral functions after a quench from $\lambda=0$ to $2$ (averaged over one phonon oscillation period and measured at $t=30$). The time average quickly approaches a quasi-steady result. The initial temperature is $\beta=5$. Top right panel: equilibrium spectral functions for $\lambda=2$ and $\beta=5$. Bottom panel: comparison of the $U=12$ time-averaged nonequilibrium spectral function (black line) to thermal spectral functions at elevated temperatures.   
}
\label{spectrum_exp}
\end{center}
\end{figure}     

Because of the periodic driving, the time-dependent spectral function obtained from the Fourier transform of the retarded Green's function is oscillating and not necessarily positive. However, the time-averaged spectral function (averaged over one phonon oscillation period) rapidly converges to the positive function shown (for $t=30$) in the top left panel of Fig.~\ref{spectrum_exp}. The overall shape is similar to the thermal spectral function at $\lambda=2$ (top right panel), but there are important differences. In particular, the time-averaged nonequilibrium spectral function cannot be reproduced by increasing the temperature of the thermal spectrum, as is illustrated for $U=12$ in the lower panel of the figure. While raising the temperature leads to a broadening of the peaks, similar to what is seen in the nonequilibrium result, the heating leads to a filling-in of the gap by more and more prominent side-bands. The time-averaged nonequilibrium spectral function, on the other hand, contains even less weight in the gap region than the original $\beta=5$ thermal spectrum. Therefore, the nonequilibrium spectral function is not similar to that of a thermal state at higher temperature, but rather resembles a broadened version of the spectral function of the initial state. 

Extracting the gap size from the peak-to-peak distance between the first prominent side-peaks in the time-averaged spectrum, we find that $U=9.25$ and $10.25$ correspond to a gap of approximately $\omega_0$ and $2\omega_0$. A plot of the doublon production rate as a function of gap size (lower panel of Fig.~\ref{slope_exp}) shows that the production of doublons is enhanced whenever the gap-size is a multiple of the phonon frequency.

\begin{figure}[t]
\begin{center}
\includegraphics[angle=-90, width=0.49\textwidth]{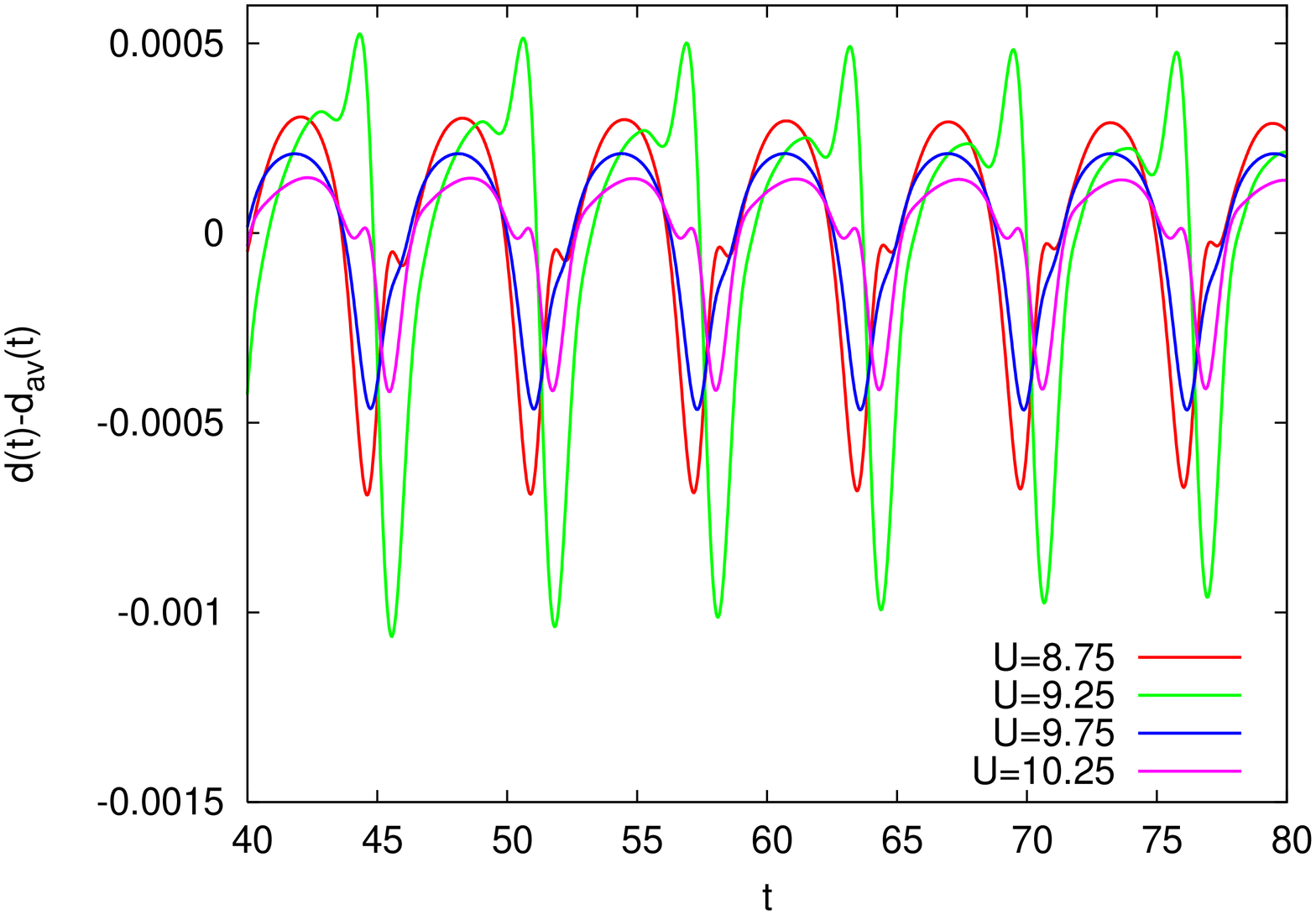}\\
\includegraphics[angle=-90, width=0.49\columnwidth]{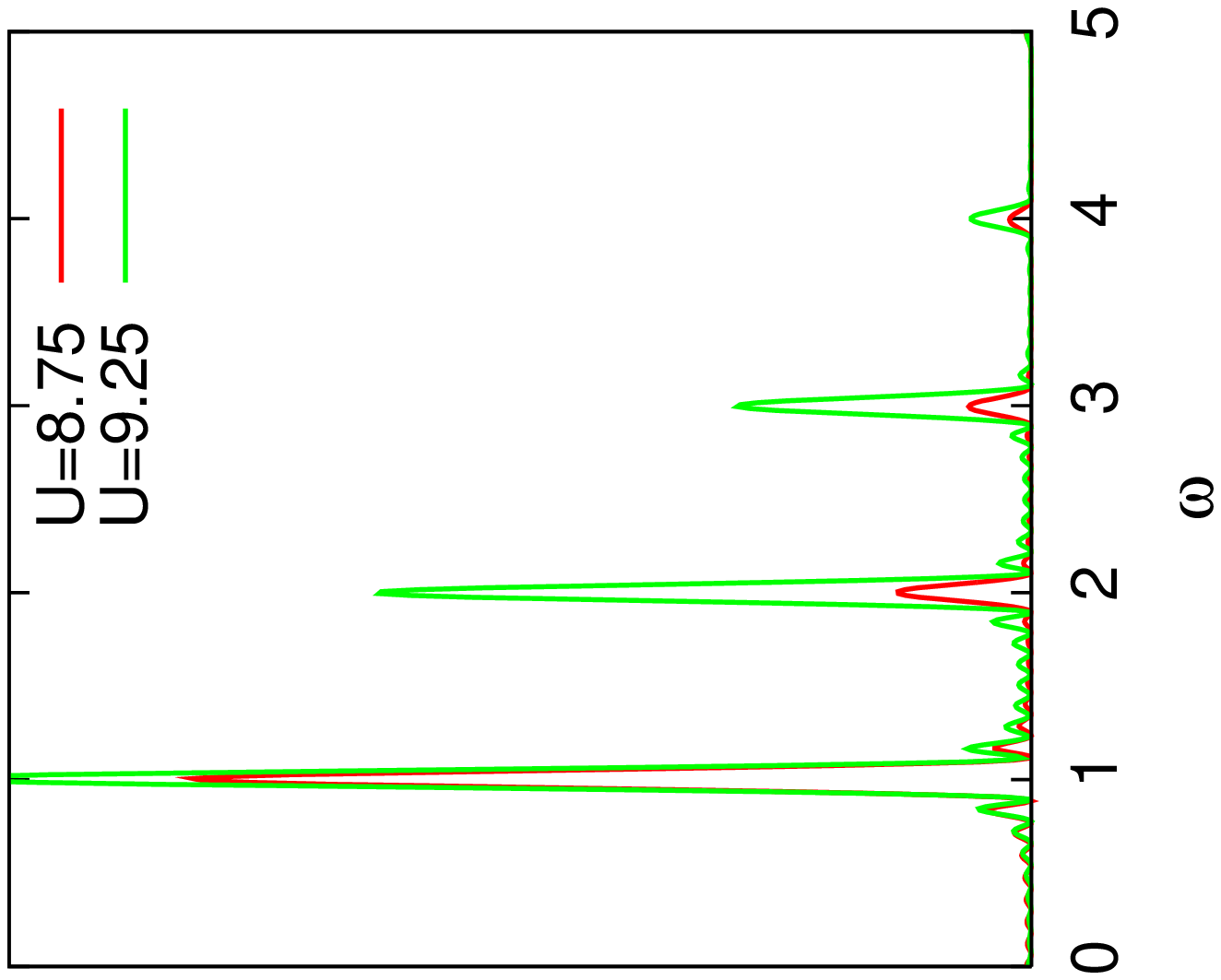}
\includegraphics[angle=-90, width=0.49\columnwidth]{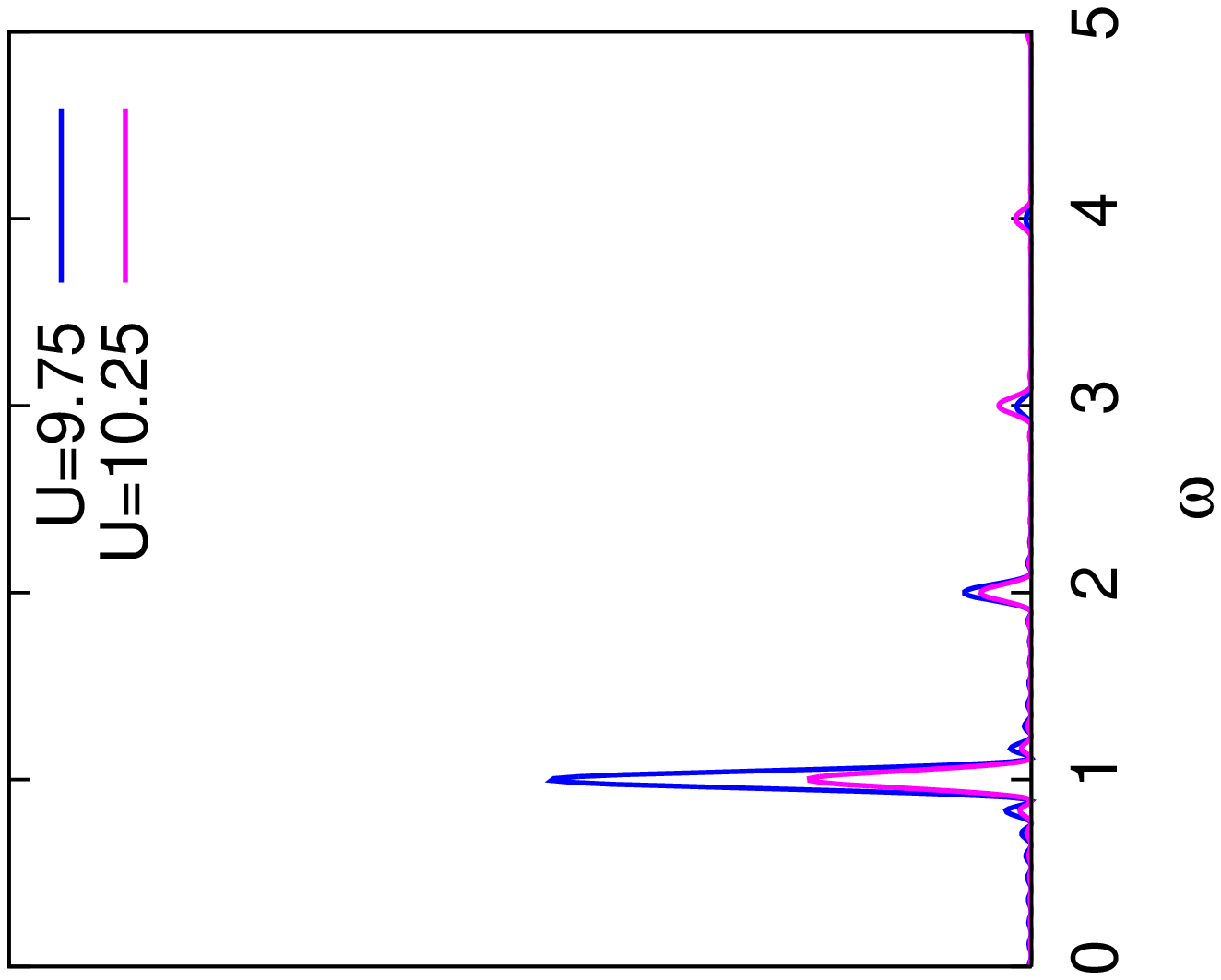}\\
\caption{
Top panel: Time-evolution of the double occupancy with time average (over one phonon oscillation period) subtracted, $d(t)-d_\text{av}(t)$, after a quench from $\lambda=0$ to $2$. Bottom panels: Fourier transforms of $d(t)-d_\text{av}(t)$ computed on the time-interval $40\le t \le 80$.
}
\label{fig_double_osc}
\end{center}
\end{figure}

Finally, let us take a closer look at the shape of the double occupancy $d(t)$ plotted in Fig.~\ref{fig_double_exp}. The zig-zag shape of some of the curves indicates that many Fourier modes are excited. In the top panel of Fig.~\ref{fig_double_osc} we subtract the time-average of $d(t)$ over one period, $d_\text{av}(t)=\frac{\omega_0}{2\pi}\int_{t-\frac{\pi}{\omega_0}}^{t+\frac{\pi}{\omega_0}}d\bar t d(\bar t)$, to extract the superimposed modulations. The Fourier transformation of $d(t)-d_\text{av}(t)$ on the time-interval $40\le t \le 80$ gives the spectra shown in the lower panel of the figure. These spectra show that the modulations are a superposition of modes with $\omega=n\omega_0$, $n=\pm 1, 2, \ldots$, and that the curve for $U=9.25$ (corresponding to phonon-enhanced doublon production) has larger contributions from higher frequency modes than the curve for $U=8.75$ (which corresponds to a minimum in the doublon production rate). A similar result is found if we compare the spectra for $U=10.25$ (maximum in the doublon production rate) and $U=9.75$ (minimum in the doublon production rate). 

\section{Summary}
\label{summary}

We have developed a formalism to treat electron-phonon couplings of the Holstein-type within nonequilibrium DMFT. A generalized Lang-Firsov transformation, based on a simultaneous (time-dependent) shift of the phonon coordiante and momentum, allows to decouple the electrons and phonons and to evaluate the phonon contribution in a strong-coupling (hybridization) expansion on the Kadanoff-Baym contour. While the resulting formalism is exact, an efficient nonequilibrium DMFT calculation requires some approximate impurity solver. We proposed  approximate schemes based on the non-crossing and one-crossing approximation, and tested them against the exact results in equilibrium. These tests suggest that even the simplest non-crossing approximation allows a qualitatively correct description of the competition between the instantaneous Coulomb repulsion and the phonon-mediated attractive interaction in the Mott insulating phase, as well as the transitions to the bipolaronic insulating phase. 

The formalism for externally driven phonons showed that in a homogeneous bulk system, the perturbation does not propagate into the DMFT self-consistency, so that the effect of the external driving is essentially trivial. Interesting effects may show up in an inhomogeneous set-up or in a two-sublattice system with a sublattice-dependent force. The investigation of these effects is left for a future project. Here, we focused on the evolution of doublons in the Mott insulating Holstein-Hubbard model, after an interaction pulse and after a rapid increase in the electron-phonon coupling. The interaction pulse provides a convenient way to excite electrons across the Mott gap (production of doublon-holon pairs in a broad spectral range) and we studied the decay of these nonthermal doublons as a function of the interaction strength. We showed that the relaxation time decreases whenever the gap in the spectral function is a multiple of the phonon-frequency. In this case doublons and holons can efficiently recombine by transferring their energy to the lattice. 

A rapid increase of the electron-phonon coupling leads to a decrease in the effective instantaneous interaction and thus to an increase of the equilibrium density of doublons. In this case one finds an enhancement of the doublon production rate whenever the gap is a multiple of the phonon frequencies. The dynamics is also strongly influenced by the excitation of the phonons during the quench, which leads to a persistent (weakly damped) oscillation in the effective electron-electron interaction and a periodic flow of energy between the electronic system and the lattice. This periodic driving leads to an essentially linear increase in the number of doublons, similar to what is observed in a Hubbard model with periodically modulated interaction.   

\acknowledgements

We thank N. Tsuji, T. Oka and H. Aoki for stimulating discussions. Some part of this work was carried out at the Aspen Center for Physics during the summer 2013 program on ``Disorder, Dynamics, Frustration and Topology in Quantum Condensed Matter". The simulations were done on the UniFR cluster. PW was supported by FP7/ERC starting grant No. 278023.

\end{document}